\documentclass[acmlarge]{acmart}

\usepackage{float}
\usepackage{booktabs}
\usepackage{longtable}
\usepackage{array}
\usepackage{url}
\usepackage{hyperref}
\usepackage{graphicx}
\usepackage{enumitem}
\usepackage{rotating}
\usepackage{tablefootnote}
\usepackage{xcolor}
\usepackage{framed}
\usepackage{subcaption}
\usepackage{tabularx}
\usepackage{eurosym}

\makeatletter
\newcommand{\myconfshort}{\acmConference@shortname}
\newcommand{\myconffull}{\acmConference@name}
\newcommand{\myconfdate}{\acmConference@date}
\newcommand{\myconfloc}{\acmConference@venue}
\AtBeginDocument{
  \fancypagestyle{firstpagestyle}{
    \fancyhead{}%
    \fancyfoot[C]{}%
  }
  \fancyhf{}
  \fancyhead[LO]{\@headfootfont\shorttitle}%
  \fancyhead[RE]{\@headfootfont\@shortauthors}%
  \fancyhead[LE]{\@headfootfont\footnotesize \myconfshort, \myconfdate, \myconfloc}%
  \fancyhead[RO]{\@headfootfont\footnotesize \myconfshort, \myconfdate, \myconfloc}%
  \fancyfoot[C]{}%
}
\makeatother
\acmBooktitle{\conffull\@ (\confshort), \confdate, \confloc}

\copyrightyear{2026}
\acmYear{2026}
\setcopyright{cc}
\setcctype{by}
\acmConference[FAccT '26]{The 2026 ACM Conference on Fairness, Accountability, and Transparency}{June 25--28, 2026}{Montreal, QC, Canada}
\acmBooktitle{The 2026 ACM Conference on Fairness, Accountability, and Transparency (FAccT '26), June 25--28, 2026, Montreal, QC, Canada}
\acmDOI{10.1145/3805689.3812301}
\acmISBN{979-8-4007-2596-8/2026/06}

\begin{document}
\title[Supporting Illegal Content Reporting under the Digital Services Act with LLMs]{AI at the Front Lines of Platform Governance: Using LLMs to Support Illegal Content Reporting under the Digital Services Act}


\author{Marie-Therese Sekwenz}
\authornote{Authors contributed equally to this research.}
\email{M.T.Sekwenz@tudelft.nl}
\affiliation{%
 \institution{Delft University of Technology, Delft}
 \city{Delft}
 \country{The Netherlands}
}
\author{Shreyan Biswas}
\authornotemark[1]
\email{S.Biswas@tudelft.nl}
\affiliation{%
 \institution{Delft University of Technology}
 \city{Delft}
 \country{The Netherlands}
}
\author{Rita Hermann-Gsenger}
\email{rita.gsenger@weizenbaum-institut.de}
\affiliation{%
 \institution{Weizenbaum Institute}
 \city{Berlin}
 \country{Germany}
}
\author{Ujwal Gadiraju}
\email{U.K.Gadiraju@tudelft.nl}
\affiliation{%
 \institution{Delft University of Technology}
 \city{Delft}
 \country{The Netherlands}
}









\newcommand{\MTS}[1]{\textcolor{black}{#1}}
\newcommand{\UG}[1]{\textcolor{black}{#1}}
\newcommand{\RG}[1]{\textcolor{black}{#1}}
\newcommand{\SB}[1]{\textcolor{black}{#1}}

\newcommand{\xai}{\textsc{\textbf{XAI}}}
\newcommand{\eai}{\textsc{\textbf{EvalAI}}}
\newcommand{\bl}{\textsc{\textbf{Baseline}}}
\begin{abstract}
Illegal content reporting mechanisms are a key technical and organizational measure through which online platforms address the dissemination of illegal content under European Union law. Under the Digital Services Act (DSA), user notices submitted pursuant to Article~16 must be sufficiently substantiated and provided in good faith, requiring users to interpret legal and procedural language and translate it into legally meaningful categories and reasons. In practice, however, reporting illegal content remains cumbersome across major social media platforms, placing substantial cognitive and legal demands on users. Without effective support at the reporting interface, operationalizing Article~16 in practice remains challenging.
We investigate how large language model (LLM)–based assistants can support illegal content reporting. In a controlled user study ($N=450$) using an interface modeled on a major platform’s reporting workflow, we compare three conditions: (1) a conventional explainable AI assistant (\xai) that suggests a single legal category with a rationale, (2) an evaluative AI assistant (\eai) that presents balanced pro and con arguments across candidate legal provisions for user deliberation, and (3) a baseline reflecting unaided reporting (\bl). We further examine these assistance forms under systematically varied AI error regimes.
Our results show that \eai{} improves provision-level accuracy under AI error regimes and reduces misclassification distance relative to conventional \xai{}, particularly for near-miss and overbreadth errors. In contrast, conventional \xai{} does not improve—and can degrade—the quality of users’ rationales relative to unaided reporting, despite enabling faster decisions when the AI output is correct. We discuss implications for the design of compliance-oriented reporting interfaces, highlighting trade-offs between accuracy, deliberation, and vulnerability to misleading AI output.

\end{abstract}





\maketitle

\section{Introduction and Background}

Online platforms have become core societal infrastructure~\cite{plantin_infrastructure_2018} and \emph{de facto} public spaces shaping political discourse, cultural participation, and access to information~\cite{european_commission_working_2020}. As a result, governments increasingly regulate how platforms address harms such as the dissemination of illegal content~\cite{clune_content_2024}. The European Digital Services Act (DSA) harmonizes content governance rules across the European Union~\cite{noauthor_regulation_2022}, aiming to create a ``safe, predictable and trustworthy online environment'' (Rec.~3 DSA).

Article~16 mandates that platforms provide user-facing mechanisms for reporting illegal content\MTS{. What is `illegal' is defined in Article 3~(h) DSA}. \footnote{ \MTS{The definition of `illegal content' states: ``any information that, in itself or in relation to an activity, including the sale of products or the provision of services, is not in compliance with Union law or the law of any Member State which is in compliance with Union law, irrespective of the precise subject matter or nature of that law." Recital~12 further clarifies: ``‘illegal content’ should broadly reflect the existing rules in the offline environment. In particular, the concept of ‘illegal content’ should be defined broadly to cover information relating to illegal content, products, services and activities. In particular, that concept should be understood to refer to information, irrespective of its form, that under the applicable law is either itself illegal, such as illegal hate speech or terrorist content and unlawful discriminatory content, or that the applicable rules render illegal in view of the fact that it relates to illegal activities."}} \MTS{The requirements of Article 16 DSA are relevant to a wide range of platforms including \emph{hosting intermediary services} (Art. 3(g) DSA), \emph{online platforms} (Art.~3(i) DSA), \emph{online search engines} (Art.3~(j) DSA), and \emph{Very Large Online Platforms} (VLOPs) and \emph{Very Large Search Engines} (VLOSEs) according to Article~33 DSA.\footnote{The VLOPs and VLOSEs have additional reporting and assessment duties compared to other platforms within the scope of the DSA. Such platforms have e.g., additional transparency reporting duties (Art.~42 DSA), or the obligation to conduct \emph{Systemic Risk Assessments} (Art.~34 DSA) and be externally evaluated through \emph{Independent Audits} (Art.~37 DSA). To qualify as a VLOP/VLOSE platforms must reach a threshold linked to "number of average monthly active recipients of the service in the Union equal to or higher than 45 million" according to Article 33~(1) DSA.}} Reports must include (i) a sufficiently substantiated explanation of the reasons (lit.~a), (ii) a URL, (iii) identifying information of the reporting user, and (iv) a good-faith statement. While other DSA provisions impose systemic risk-management obligations \MTS{for VLOPs and VLOSEs} (Arts.~34--35,~37), Article~16 \MTS{(which is  applicable to hosting intermediary services, online platforms, search engines, VLOPs/VLOSEs)}, explicitly foregrounds the role of end users in initiating enforcement processes. Non-compliance can result in fines of up to 6\% of global annual turnover for providers of VLOPs and VLOSEs (Art.~74). Recent enforcement actions against major platforms~\cite{european_commission_commission_2025}, including a \EUR{120} {million} fine against the platform X~\cite{european_commission_commission_2025}, underscore the regulatory importance of DSA compliance. 

The DSA further expands platform responsibility to content that is \MTS{blurring the boundaries of `illegality' in cases of disinformation\footnote{See for example Art.~34(1)(c) which covers ``any actual or foreseeable negative effects on civic discourse and electoral processes." This if further discussed in Recital~84 ``[...] When assessing the systemic risks identified in this Regulation, those providers should also focus on the information that is not illegal, but contributes to the systemic risks identified in this Regulation. Such providers should therefore pay particular attention on how their services are used to disseminate or amplify misleading or deceptive content, including disinformation."} which is at times referred to as} ``lawful but awful'' \MTS{content~\cite{keller_lawful_2022, chesterman2024lawful}}. Regulating speech, therefore, entails risks of over-removal and requires trust in regulators, platform participation, and avenues for user challenge and involvement. However, accurately distinguishing illegal content from harmful but lawful expression remains difficult even for experts, let alone ordinary users~\cite{macdonald_moderating_2024}. Reporting requires mapping ambiguous posts to specific legal provisions under uncertainty. Errors at this stage can lead to under-reporting, over-reporting, or misrouting, with downstream consequences for both users and platforms. While platforms have no general monitoring duty (Art.~8), \MTS{intermediary services (Rec.~20) and hosting providers (Rec.~22), however, do not have a \emph{carte blanche} for liability questions mitigation as ``other actors [...] should also help to avoid the spread of illegal content online, in accordance with the applicable law" (Rec.~27) and ``it is important that all providers of hosting services, regardless of their size, put in place easily accessible and user-friendly notice and action mechanisms that facilitate the notification of specific items of information that the notifying party considers to be illegal content" (Rec.~50)}. \MTS{VLOPS and VLOSEs, furthermore}, are obliged to mitigate the dissemination of illegal content (Arts.~34--35\MTS{, Rec.~80 and 87}). Alternative approaches, such as community-based moderation, face challenges of scale~\cite{gillespie_content_2020,nicholas_toward_2023,douek_content_2022}, time constraints~\cite{allen_hurry_2023}, increasingly convincing AI-generated material~\cite{klincewicz_slopaganda_2025}, linguistic diversity~\cite{mehta_understanding_2025,kumar_watch_2024}, evolving semiotics, and heterogeneous content types~\cite{jiang_trade-off-centered_2023,huang_systematic_2022,zannettou_quantitative_2020}.

In response, platforms have increasingly explored AI-based support for moderation and reporting workflows~\cite{marsden_platform_2020}. Yet, AI assistance can amplify systemic risks rather than mitigate them, and human collaboration with decision-support systems is error-prone~\cite{biswas2026belief,erlei2026life}. A dominant approach to address these challenges has combined decision support with explainable AI (XAI)~\cite{bertrand_how_2022,schemmer2022meta,he2024opening}, aiming to foster appropriate reliance through post-hoc rationales~\cite{schemmer2023appropriate,gadiraju2025enterprising}. This paradigm assumes relatively reliable recommendations and treats explanations as justificatory aids. In legal classification tasks, however, users must deliberate among competing interpretations \emph{before} committing to a category, and a single suggested label may anchor decisions even when incorrect~\cite{nourani2021anchoring}. A growing body of work shows that appropriate reliance remains difficult to achieve—even with XAI—and that explanations can increase overreliance via illusory explanatory depth, particularly in low-friction or conversational settings or when explanations are hard to verify~\cite{schemmer2022should,kuper2025psychological,he2024err,mehrotra2025even,salimzadeh2024dealing,he_knowing_2023,ehsan2024seamful,kaur2022sensible,he_is_2025,fok2024search}.

To address these limitations, \citet{miller2023explainable} proposed \emph{evaluative AI}: decision support that provides evidence for and against multiple options rather than prescribing a recommendation. This hypothesis-driven approach surfaces structured arguments, statutory elements, and evidentiary considerations without prescribing an outcome, thereby supporting user deliberation and mitigating over- and under-reliance. Evaluative AI is particularly suited to legal reporting contexts, where responsibility rests with the user and overconfident automation may undermine agency and legal reasoning.

In this work, we examine how different architectures of AI assistance influence provision-level accuracy, the quality of users' substantiated explanations, and robustness under erroneous AI output in illegal-content reporting. We do not frame AI assistance as a mechanism for automated compliance or enforcement. Instead, we focus on user--AI interaction at the reporting interface and its implications for users' ability to meet the procedural requirements of Article~16 DSA \MTS{relevant to all intermediary services}. References to \MTS{VLOP/VLOSE's} systemic-risk provisions are limited to downstream effects of misclassification at the reporting stage, rather than claims of direct compliance or risk-mitigation effects (Arts.~34--35).

\SB{To that extent, our study is guided by the following overarching research question---\emph{``How do different AI assistance architectures in illegal-content reporting interfaces shape user decision quality, explanation quality, and robustness to erroneous AI output in Article~16 DSA reporting tasks?''}} %
\SB{Based on this question and informed by existing literature, we derive six hypotheses that henceforth structure our empirical analysis. As described below, H1--H3 examine provision-level accuracy under correct and erroneous AI assistance. H4 examines whether assistance architecture improves the quality of users' own substantiated explanations. H5 tests whether different AI error conditions differentially degrade performance and whether \eai{} attenuates these losses relative to \xai. H6 examines downstream risk-relevant outcomes of misclassification at the reporting stage, namely over-removal and misrouting.}

\paragraph{Hypotheses}
\begin{framed}
\begin{itemize}[leftmargin=*]
    \item \SB{\textbf{H1}: Under AI error conditions, participants using \eai{} will achieve higher provision-level accuracy than participants using \xai{}, indicating lower overreliance on incorrect AI outputs.}

    \item \SB{\textbf{H2}: Under AI error conditions, participants using \eai{} will select legally closer incorrect provisions than participants using \xai{} (smaller misclassification distance), but this reduction will be attenuated for Out-of-Scope errors relative to Near-Miss and Overbreadth errors.}

    \item \SB{\textbf{H3}: When AI assistance is correct, \eai{} and \xai{} will yield comparable provision-level accuracy, but decisions under \xai{} will be faster than under \eai.}

    \item \SB{\textbf{H4}: Participants using \eai{} will produce higher-quality substantiated explanations than those in \bl{} and \xai{}, as reflected in stronger legal element coverage and reasoning depth.}

    \item \SB{\textbf{H5}: AI error conditions (Near-Miss, Overbreadth, Out-of-Scope) will differ in the magnitude of their impact on reporting performance, and \eai{} will differentially attenuate these error-specific performance drops relative to \xai.}

    \item \SB{\textbf{H6}: Relative to \xai{}, \eai{} will better align with downstream DSA-relevant reporting goals by reducing over-removal under Overbreadth errors and reducing misrouting under Out-of-Scope errors.}
\end{itemize}
\end{framed}

We investigate these hypotheses through a controlled experiment ($N=450$) using a custom reporting interface modeled on existing social-media workflows. Participants evaluated a simulated feed containing illegal content grounded in German criminal law,\footnote{We selected this context as a member-state example where illegal-content reporting was regulated prior to the DSA~\cite{wagner_regulating_2020}.} selected mandatory legal categories, and submitted reports under one of three assistance conditions. The primary outcomes in our study capture provision-level legal accuracy; while secondary outcomes assess explanation quality, decision time, and trust-related measures.

This paper contributes to FAccT scholarship on content moderation and platform governance in three ways:
\begin{enumerate}[leftmargin=*]
    \item introducing an experimental paradigm for AI-assisted content reporting under regulatory constraints;
    \item providing empirical evidence on how assistance conditions affect accuracy, reasoning quality, decision time, and trust under correct and erroneous AI assistance; and
    \item deriving design and policy implications for DSA-aligned reporting interfaces.
\end{enumerate}

\section{Related Work}
\SB{Prior research examines how generative AI tools support human tasks across domains such as design \cite{govindankutty_design_2024, shin_paper_2024}, finance \cite{peta_user-friendly_2024}, creative writing \cite{biswas_mind_2025}, and content moderation \cite{ozanne_shall_2022}. Across these settings, effectiveness depends not only on model quality, but also on how assistance is presented, how easily outputs can be verified, and whether the interface supports appropriate reliance rather than passive acceptance \cite{he_plan-then-execute_2025,deng2025proactive,subramanian2023uncertainty}.}

\textit{\textbf{Content Moderation and Reporting}.}
User reporting plays a central role in content moderation ecosystems \cite{wei_theres_2023}, yet the scale and complexity of online platforms exceed what purely human moderation can sustain \cite{marsden_platform_2020}. As a result, platforms rely on automated moderation systems that classify user-generated content and trigger governance outcomes \cite{gorwa_algorithmic_2020}. Hybrid approaches that combine automation with user input show promise, but also raise concerns about overreliance on AI \cite{he_knowing_2023, huang_survey_2025}. 

Content moderation is often described as operating across two phases: \emph{ex ante} moderation before publication, typically relying on automated filtering, and \emph{ex post} moderation after publication, where user reporting plays a key role  \cite{gillespie_content_2020}.  Reporting interfaces, however, are often opaque and misaligned with user expectations \cite{kou_flag_2021} and mental models \cite{zhang_cleaning_2024}. Users are reluctant to engage deeply with reporting \cite{jhaver_personalizing_2023}, though they value nuanced options, proactive platform action, and personal moderation settings \cite{jhaver_users_2025}. User characteristics further shape flagging behavior \cite{naab_flagging_2018}. While accessible platform help centers can aid policy understanding \cite{mayworm_online_2024}, most users lack the legal knowledge to file substantiated reports \cite{sekwenz_it_2025}, echoing broader findings on the inaccessibility of rules like the terms of service \cite{mcdonald_cost_2008,obar_biggest_2020}.

\noindent\textit{\textbf{Legal Design Challenges}.}
The DSA highlights legal design (Art.~25) and bans features that ``deceive or manipulate [...] or otherwise materially distort or impair" user decision-making \cite{di_porto_collective_2023, leiser_dark_2023, herman_dark_2024} \MTS{as exemplified in the Commission's first DSA-based fine against X for the ``use of the ‘blue checkmark' for ‘verified accounts' [which] deceives users." \cite{european_commission_commission_2025}}. The DSA also requires ``user-friendly" reporting mechanisms (Art.~16). Yet, which design solutions meet these standards remains unclear \cite{sekwenz_it_2025}. 
As regulation reshapes design practice, interdisciplinary approaches become necessary \cite{gray_legal_2024}. Poorly implemented legal UX can take the form of \emph{dark patterns}, raising usability, legal, and ethical concerns \cite{gray_dark_2018} under which reporting mechanisms as \emph{Interface Interference} can be subsumed, driven by \emph{Language Inaccessibility} and \emph{Complex Language} \cite{gray_ontology_2024}.

\noindent\textit{\textbf{AI-supported reporting}.}
Large Language Models (LLMs) as a form of AI, increasingly shape moderation systems \cite{gomez_algorithmic_2024}, including detecting harmful content across languages \cite{nicholas_toward_2023,kumar_watch_2024} and generating ``Statements of Reason" under the DSA \cite{aspromonte_llms_2024}. LLMs are also applied in legal practice and research to simplify statutes and provide advice \cite{hacker2020explainable,lai2024llmlaw,ogunde2024generative}, but they remain prone to hallucinations and misclassifications \cite{huang_survey_2025,leiser_hill_2024}, raising ethical concerns. 

\SB{\citet{salimzadeh_dealing_2024} show that complexity and uncertainty shape user performance in AI-supported tasks; in legal content reporting, this makes meaningful contestability especially important because interfaces must enable users to understand, challenge, and revise automated reasoning \cite{yurrita_identifying_2025,cobbe2021reviewable,kingston2017ai_gdpr}.}

\SB{A central challenge in AI-supported reporting is not only whether AI assistance is present, but how it is structured and how susceptible users become to inevitable output errors. Conventional explainability often follows a \emph{recommend-and-justify} logic, foregrounding a single predicted option and rationale. Such designs may reduce friction, but can also anchor users to erroneous outputs when the reasoning is difficult to verify \cite{buccinca2021trust,vasconcelos2023explanations,he_is_2025,gupta_trust_2022,mehrotra_even_2025,kostric_should_2025,kocielnik_will_2019,jeung_correct_2023}. By contrast, more deliberative interfaces may reduce harmful overreliance, though often at a cost in effort and usability \cite{buccinca2021trust,miller2023explainable,vasconcelos2023explanations}}.
\SB{Taken together, this literature suggests that assistance architectures foregrounding a single recommendation may be especially vulnerable to harmful reliance under AI error, whereas evaluative architectures that present evidence for and against alternatives may better align with human decision processes and support correction and recovery. This expectation informs \emph{H1} and the premise of \emph{H2}, which test whether Evaluative AI improves provision-level accuracy and reduces misclassification distance relative to Conventional XAI when the AI advice is wrong.}

\SB{Recent work proposes more argumentative and contestable forms of AI support \cite{miller2023explainable,freedman2025argumentativelargelanguagemodels,yurrita_identifying_2025,yurrita2025towards}. User judgments are also sensitive to how alternatives are presented under model error, highlighting the importance of choice structure and choice independence \cite{erlei_understanding_2024,chang_socrasynth_2024}. Together, these lines of work point toward \emph{evaluative} forms of AI assistance that support deliberation across alternatives rather than merely justifying one recommendation. This informs \emph{H3}, which examines whether the two AI-assisted conditions remain comparable in accuracy when the AI advice is correct, while differing in decision time.}

\SB{Because evaluative interfaces expose users to structured supporting and opposing considerations, they may support richer user justifications as well as better selections \cite{miller2023explainable,freedman2025argumentativelargelanguagemodels}, motivating \emph{H4}. Prior work on automation bias, error tolerance, and contestability also suggests that plausible, overbroad, and legally irrelevant errors may shape reliance differently and carry distinct downstream consequences \cite{erlei_understanding_2024,buccinca2021trust}. This motivates \emph{H5}, which tests whether AI error conditions differ in their performance impact and whether Evaluative AI attenuates those penalties relative to Conventional XAI. It also motivates \emph{H6}, which examines whether these differences extend to downstream risk-relevant outcomes such as disproportionate intervention under Overbreadth errors and routing failures under Out-of-Scope errors (described in detail in the following section).}

\SB{Together, these strands of work motivate our comparison of manual reporting, conventional XAI, and evaluative AI under systematically controlled legal error conditions in a user-facing illegal-content reporting task.}
\section{Experimental Design and Procedure}

We conducted a controlled experiment with a between-subjects factor (\emph{assistance condition}) and a micro within-subject manipulation (\emph{correct AI assistance} vs.\ \emph{AI error condition}) applied only in the AI-assisted arms. Participants were blind to their condition assignment.

\paragraph{Assistance conditions}
Participants were randomly assigned to one of three conditions:
\begin{itemize}
    \item \textbf{\bl}: Manual reporting without AI assistance (control).
    \item \textbf{\eai}: Evaluative AI, presenting structured pros and cons for multiple candidate legal provisions, inspired by \citet{miller2023explainable}.
    \item \textbf{\xai}: Conventional XAI, presenting a single suggested legal provision with a brief rationale.
\end{itemize}

\begin{figure}[ht]
\centering
\begin{subfigure}[c]{0.28\linewidth}
    \centering
    \includegraphics[width=\linewidth,trim=20 40 20 40,clip]{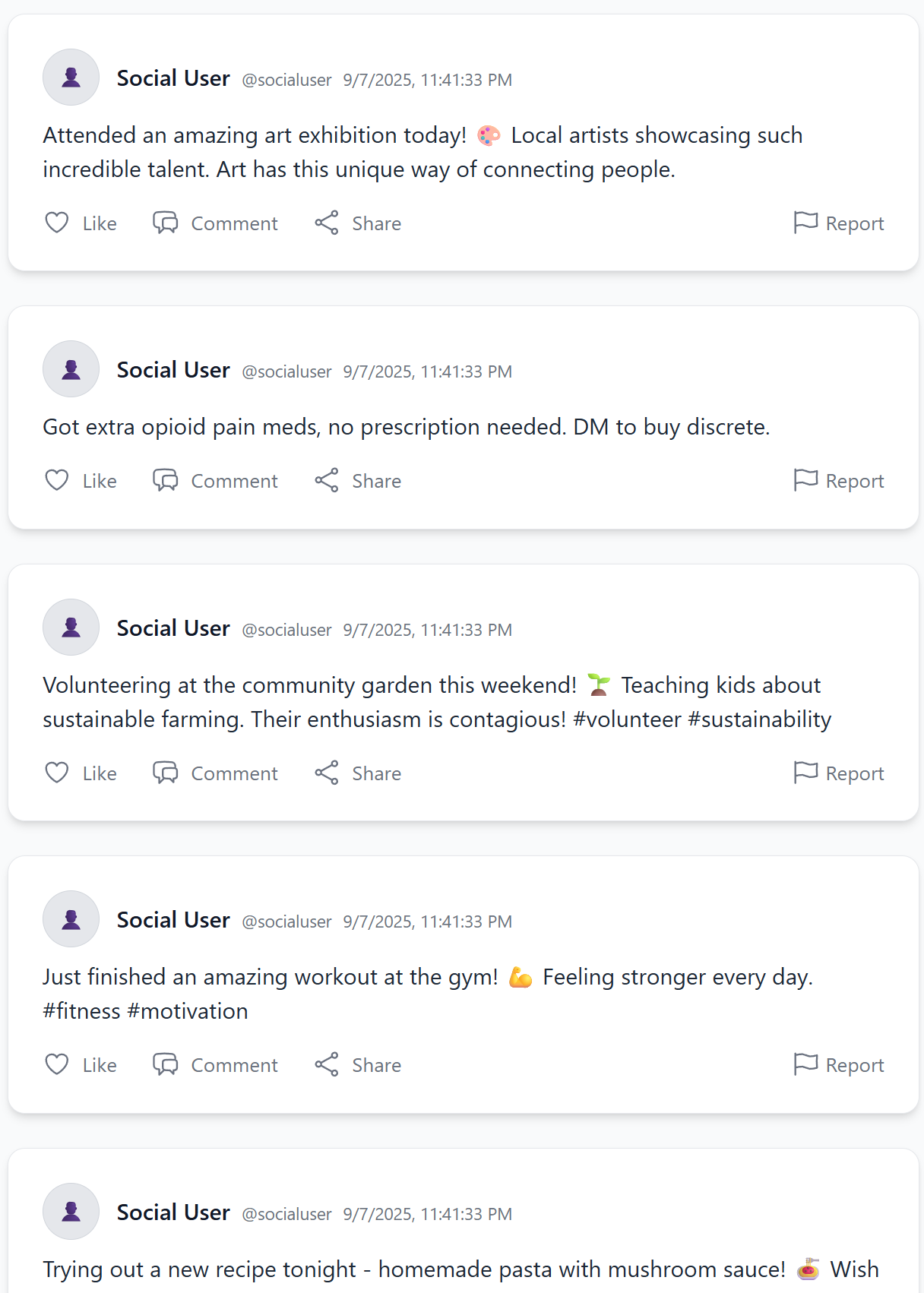}
    \caption{The social media feed}
\end{subfigure}
\hfill
\begin{subfigure}[t]{0.68\linewidth}
    \centering
    \includegraphics[width=\linewidth,trim=40 120 40 80,clip]{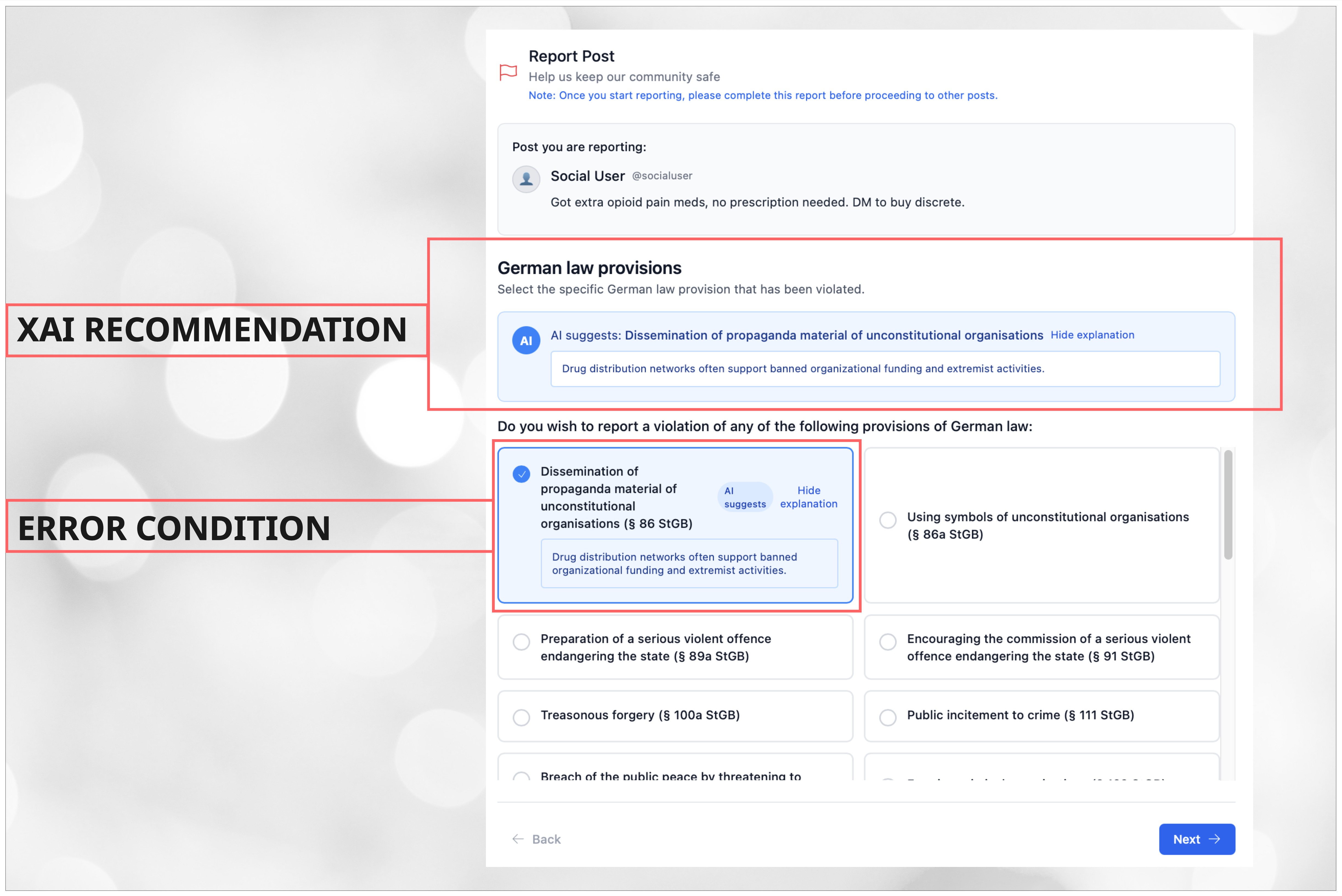}
    \caption{Evaluative AI showing pro/con arguments for a hate speech case.}
    \vspace{0.4em}
    \includegraphics[width=\linewidth,trim=40 120 40 80,clip]{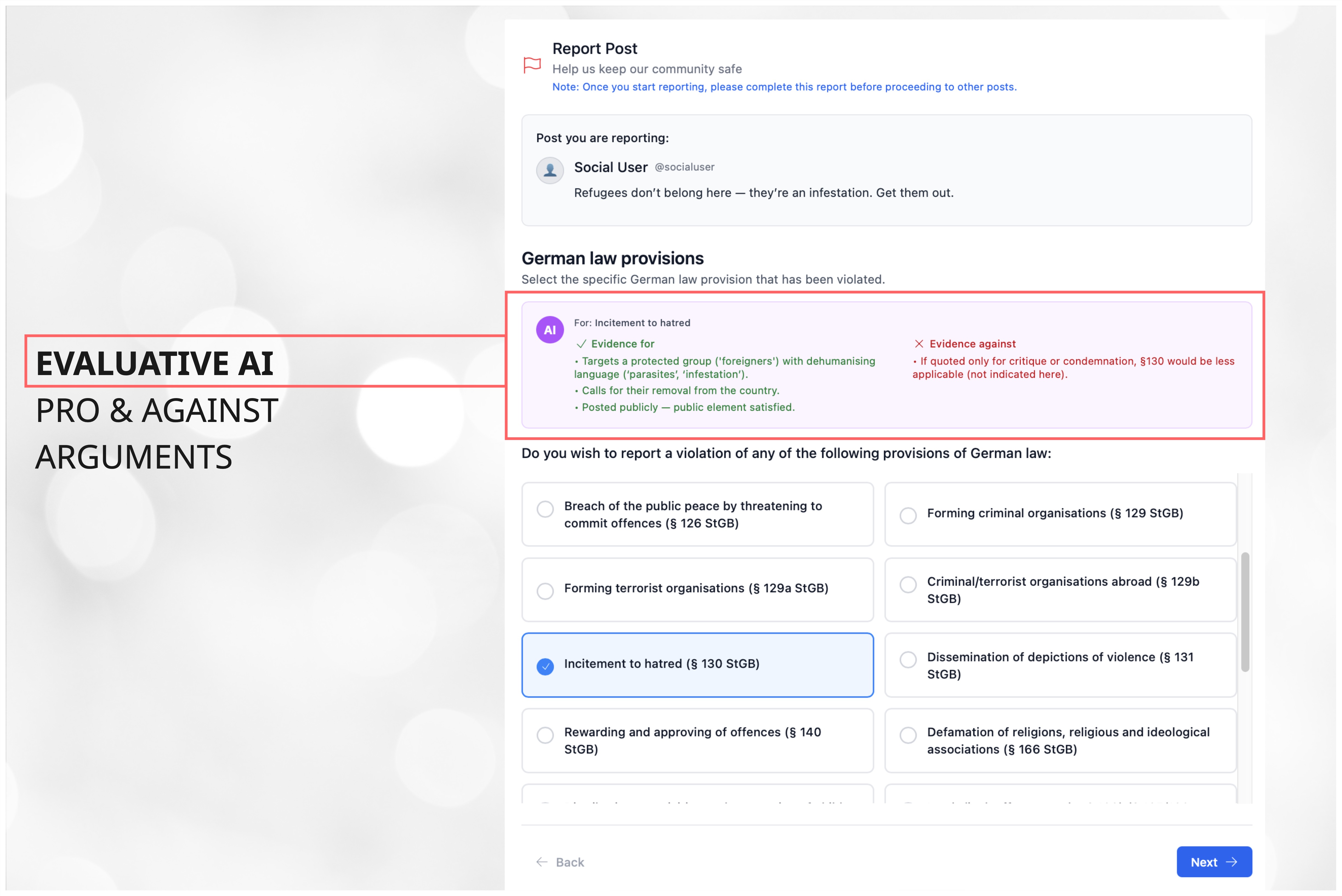}
    \caption{Example of XAI recommendation error: the system suggests an incorrect legal provision.}
\end{subfigure}
\caption{Components of the experimental interface. (a) the social media feed the users were presented }
\Description{The figure shows three screenshots of the experimental reporting interface. Panel A shows a simulated social media feed with several short posts and standard interaction buttons. Panel B shows the evaluative AI condition, where participants can inspect structured pro and con arguments for candidate German legal provisions in a hate speech reporting task. Panel C shows the conventional XAI condition in an error scenario, where the interface highlights an incorrect legal provision as the AI recommendation and provides a short rationale for that recommendation.}
\label{fig:interface_conditions}
\end{figure}

Each participant evaluated a 12-post feed~\ref{fig:interface_conditions}: 10 benign filler posts (identical across participants) and two illegal posts drawn from four legal case categories (\S~86a symbols [L\_6], \S~111 incitement [L\_10], \S~130 hate speech [L\_15], drug sales [L\_27]; Appendix~\ref{tab:cases-appendix}). The two illegal posts were A/B variants of the same case. In AI-assisted conditions, one illegal post included correct AI assistance and the other included an injected AI error condition. Error order was counterbalanced.

\SB{\paragraph{AI assistance architectures}
Figure~\ref{fig:interface_conditions} illustrates the reporting interface in the two AI-assisted conditions. In all conditions, participants saw the same candidate legal provisions. What varied across conditions was the {assistance architecture}. In \xai{}, the interface pre-highlighted one AI-suggested provision and displayed a short rationale (\emph{recommend-and-justify}). In \eai{}, no provision was preselected; instead, participants could inspect structured supporting and opposing arguments for candidate provisions (\emph{deliberation-oriented}). In \bl{}, no AI-generated reasoning was shown.}

\paragraph{AI error conditions}
\SB{To capture distinct human--AI risk profiles, we introduced three randomized AI error conditions that meaningfully operationalize errors in our context:}
\begin{itemize}
    \item \textbf{Near-Miss (NM):} A semantically close but incorrect provision (e.g., \S~111 instead of \S~130), representing a setting in which the AI output is plausible and may increase automation bias risk~\cite{mosier2018human,dzindolet2002perceived}.
    \item \textbf{Overbreadth (OB):} A related but overly broad provision (e.g., \S~166 instead of \S~130), representing a setting that may encourage over-removal or overly expansive legal interpretation~\cite{gillespie2018custodians,gorwa_algorithmic_2020}.
    \item \textbf{Out-of-Scope (OS):} A legally irrelevant provision, representing a setting in which the AI output misroutes the report away from the appropriate legal frame~\cite{barocas2020fairness}.
\end{itemize}

\SB{These conditions are not intended to encode a universal severity ranking. Rather, they operationalize distinct reporting risks at the human--AI interface: plausible misguidance (Near-Miss), overbroad categorization (Overbreadth), and legally irrelevant routing (Out-of-Scope).}

\paragraph{Condition manifestation}
\SB{In both AI-assisted conditions, the injected error pointed to the same incorrect legal provision for a given case; only the presentation architecture differed. In \xai{}, the incorrect provision was highlighted and an explanation was provided. In \eai{} the same incorrect provision received comparatively stronger supporting arguments, while competing categories were comparatively downplayed; outside the targeted mappings, categories were often represented with weaker, minimal, or placeholder pro/con content rather than fully elaborated arguments. In \bl, no AI assistance was provided and no error manipulation was applied. Representative mappings are listed in Appendix Table~\ref{tab:error_examples}.}

\begin{figure}
    \centering
    \includegraphics[width=0.65\linewidth]{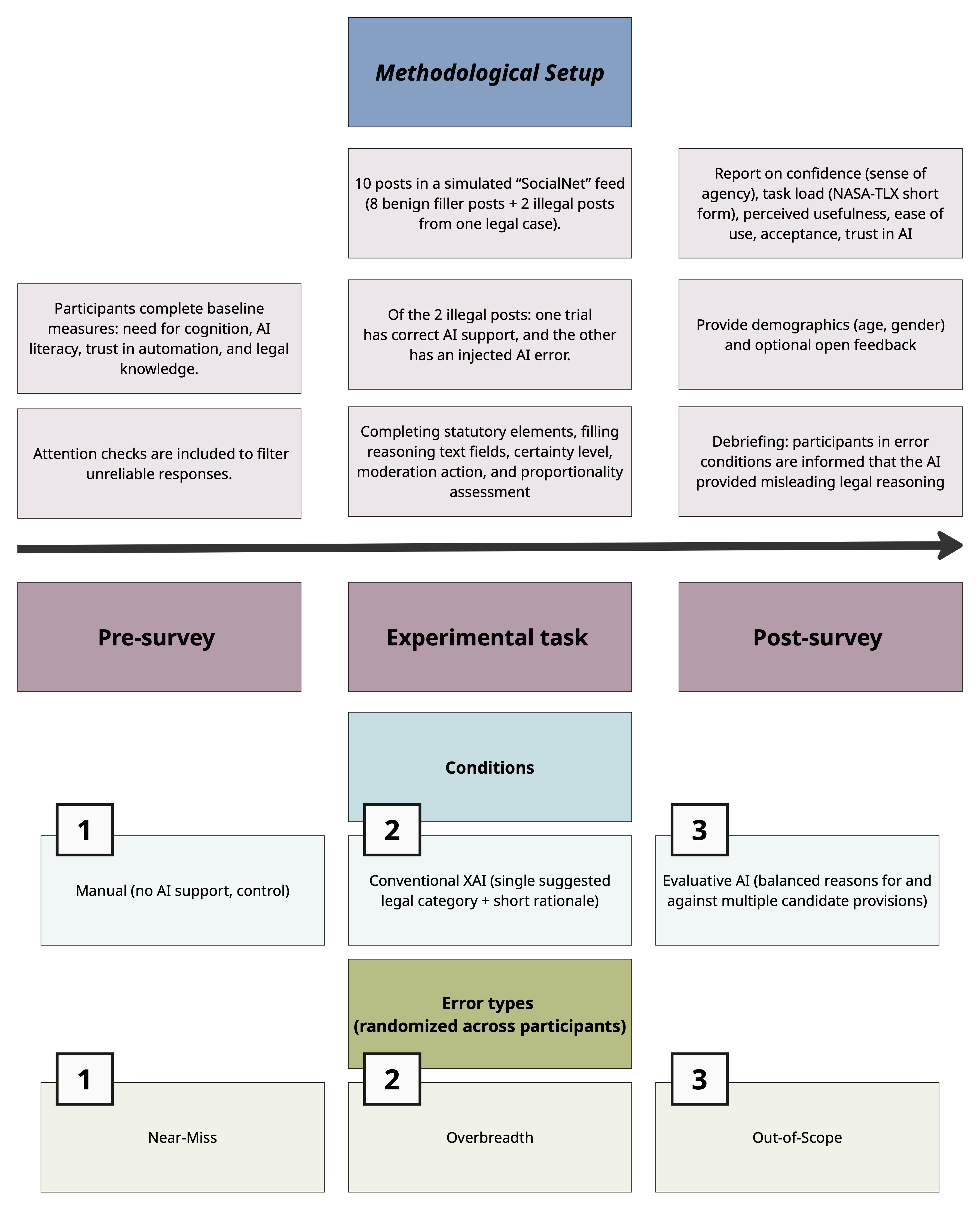}
 \caption{Overview of the experimental flow, assistance conditions, and randomized AI error conditions used in the study.}
    \label{fig:methodology_Overview}
    \Description{This figure illustrates the experimental design of the study. 
It consists of three stages: a pre-survey, an experimental task, and a post-survey. 
In the pre-survey, participants complete baseline measures such as need for cognition, AI literacy, trust in automation, and legal knowledge, with attention checks to ensure reliability. 
The experimental task is carried out in a simulated “SocialNet” feed containing 12 posts (10 benign and 2 illegal). Of the illegal posts, one is presented with correct AI support and the other with an injected AI error. Participants complete statutory elements, provide reasoning, certainty levels, and moderation actions. Three assistance conditions are tested: Manual (no AI), Conventional XAI (a single suggested legal category with rationale), and Evaluative AI (balanced reasons for and against provisions). Error types are randomized across participants. 
In the post-survey, participants report on confidence, workload, usefulness, ease of use, trust in AI, and provide demographics. Participants in error conditions are debriefed that the AI advice included misleading legal reasoning.}
\end{figure}

\paragraph{Procedure}
The study consisted of a pre-survey, the reporting task, and a post-survey (Appendix~\ref{questionnaire}--\ref{post_questionnaire}; Fig.~\ref{fig:methodology_Overview}).
\textbf{Pre-survey.} Participants completed measures of need for cognition~\cite{lins_de_holanda_coelho_very_2020}, AI literacy (MAILS)~\cite{carolus_mails_2023}, trust in automation~\cite{koerber_theoretical_2019}, and legal knowledge, including attention checks (Appendix~\ref{TiA}--\ref{Legal_Knowledge}).
\textbf{Reporting task.} Participants reviewed the 12-post feed under their assigned condition. Reporting required selecting a legal category (with AI support, where applicable), completing an \emph{Explanation of Reasons} (statutory element marking and textual evidence; Art.~16(2)(a) DSA), choosing a moderation action, and indicating proportionality. To preserve internal validity, progression was locked and cross-post comparison was prevented.
\textbf{Post-survey.} Participants completed measures of perceived usefulness and ease of use (TAM)~\cite{davis_perceived_1989}, sense of agency~\cite{tapal_sense_2017}, and workload (NASA--TLX short)~\cite{hart_development_1988}. Participants exposed to AI errors were debriefed.

\paragraph{Cases}
We modeled the reporting interface on Facebook’s illegal-content workflow using German statutory categories (Fig.~\ref{fig:legal_form_example}). Legal cases were derived from German case law, translated into fictionalized scenarios using ChatGPT (Appendix~\ref{prompting}), and cross-checked by two authors. Four cases were selected to span easy and hard classification challenges~\cite{huang_content_2025,salimzadeh_missing_2023} (See, Appendix Table~\ref{tab:posts}).

\paragraph{Variables}
The {independent variables} were assistance condition (between-subjects), AI trial type (correct-assistance vs.\ error trial; within-subjects), and AI error condition (NM, OB, OS) for error trials only.

\textit{Primary outcomes} were (a) categorization accuracy (binary match to ground truth) and (b) \SB{misclassification distance} (0 = correct; 1 = near-miss; 2 = related-but-wrong; 3 = out-of-scope).
\textit{Secondary measures} included decision time, Explanation-of-Reasons coverage, and explanation quality (1--5 rubric; two independent coders; Appendix~\ref{Codebook}). Additional measures captured moderation actions, false positives on benign posts, interaction logs, self-reported confidence, trust, workload, and demographic covariates.

\textit{Participants.}
We recruited Prolific participants located in Germany (fluent in English; approval rate $\ge 90\%$). Compensation was set to a fixed £1.60 base payment (10--15 min) in addition to a bonus up to £2.00 (i.e., a maximum of £3.60). Participants could earn bonus payments by correctly categorizing malicious posts and by providing high-quality explanations of reasons, assessed based on coverage of key legal elements and proportionality considerations. We recruited 489 participants in total, and after exclusions due to technical issues or false-positive reports on benign filler posts, $N{=}450$ were retained, with 150 participants per condition (\eai, \xai, and \bl; Appendix Table~\ref{tab:descriptives}). The final sample comprised 293 men, 152 women, and 5 non-binary participants, with mean age $M{=}31.32$ years ($SD{=}8.7$, range 18--71).

\section{Methodology}

\SB{We ran a controlled experiment following Ozanne et al.\ \cite{ozanne_shall_2022} using a custom web app (\emph{SocialNet}) that mimics a social-media reporting workflow. We evaluate six hypotheses using models matched to outcome type: logistic models for binary outcomes, ordered logit models for misclassification distance, linear models for decision time and coder-mean explanation ratings, and paired difference-in-differences specifications for error-induced performance changes. Full model equations and particulars, robustness checks, and supplementary analyses are reported in the appendix. Our goal is to test whether different assistance conditions shape reporting performance and downstream risk-relevant outcomes in DSA-aligned reporting tasks (Table~\ref{tab:hypotheses}).}

\textbf{H1: Provision accuracy under AI error.}
\SB{Among AI-assisted participants (Evaluative AI vs.\ Conventional XAI), we estimate the effect of Evaluative AI on provision-level accuracy on illegal-content error trials (one error trial per participant by design). The outcome of interest here is binary provision accuracy. We fit a logistic model with AI error condition and case fixed effects, reporting robust standard errors, odds ratios, 95\% confidence intervals, and two-sided $p$-values. Robustness checks add pre-task covariates and exclude participants who reported benign filler posts.}

\textbf{H2: Misclassification distance under AI error.}
\SB{On illegal-content error trials in the AI-assisted conditions, H2a tests whether Evaluative AI reduces misclassification distance relative to Conventional XAI, and H2b tests whether this effect varies across AI error conditions, with particular attention to Out-of-Scope errors. The outcome is an ordinal misclassification distance score ($0 {=}$ correct, $3 {=}$ farthest misclassification). We estimate proportional-odds ordered logit models with participant clustered robust standard errors. Robustness checks add pre-task covariates, exclude participants who reported benign filler posts, and compare conditions using non-parametric tests.}

\textbf{H3: No-error trials (correct AI assistance).}
\SB{We analyze the no-error trial in the two AI-assisted conditions only (one no-error trial per participant; $N{=}300$, 150 per condition). H3a tests provision-level accuracy using a binomial GLM with robust standard errors and a TOST equivalence test on the difference in proportions (\(\delta = 0.05\)). H3b tests decision time using OLS on log-transformed response times with robust standard errors, with a one-tailed Mann--Whitney U test as a non-parametric check. Robustness checks add case fixed effects where available.}

\textbf{H4: Explanation-of-Reasons quality.}
\SB{We analyze illegal-content reports with completed coder ratings across all three assistance conditions (\bl{}, \xai{}, \eai{}). Each report was independently rated by two trained annotators. We examine four coder-rated Explanation-of-Reasons dimensions: element coverage, proportionality reasoning, reasoning depth, and perceived overall quality. For each dimension, the dependent variable is the mean of the two coder ratings. We estimate linear models with (AI) assistance condition as a categorical predictor, participant-clustered robust standard errors, and pairwise contrasts across conditions. Participants could contribute up to two rated reports, which motivates clustering at the participant level. Inter-rater reliability was acceptable to high across dimensions (Appendix), supporting aggregation.}

\textbf{H5: AI error condition impact and attenuation.}
\SB{For AI-assisted participants only, we construct a paired difference-in-differences panel in which each participant contributes one illegal-content no-error trial and one illegal-content error trial. The participant’s AI error condition (Near-Miss, Overbreadth, or Out-of-Scope) is assigned at the pair level. We examine two classes of outcomes: provision accuracy and the four coder-rated Explanation-of-Reasons dimensions. Provision accuracy is analyzed using a paired difference-in-differences logistic model with participant-clustered robust standard errors. Explanation outcomes are analyzed primarily using participant-level first-difference OLS models on coder-mean ratings, with ordered-logit sensitivity analyses reported in the appendix. H5a tests whether the error penalty differs by AI error condition; H5b tests whether Evaluative AI attenuates these error-specific performance drops relative to Conventional XAI.}

\textbf{H6: DSA-relevant downstream risk outcomes.}
\SB{H6 evaluates whether, relative to Conventional XAI, Evaluative AI yields user reporting behavior more consistent with downstream risk-relevant outcomes in DSA-aligned reporting. Analyses are restricted to AI-assisted error trials and estimated separately by AI error condition. For Overbreadth errors, we model over-removal as a binary indicator of whether the participant selected a severe enforcement action. For Out-of-Scope errors, we model misrouting as a binary indicator of whether the participant failed to route the case as out-of-scope. Each outcome is analyzed with a binomial GLM with a logit link and participant-clustered standard errors. When variation is insufficient for model estimation, we report descriptive statistics only.}

\section{Results}
\SB{We first report provision-level outcomes under AI error (H1--H2), then comparisons under correct AI assistance (H3), followed by Explanation-of-Reasons outcomes (H4--H5) and downstream risk-relevant outcomes (H6).}

\begin{figure}[ht]
  \centering
  \includegraphics[width=0.8\linewidth]{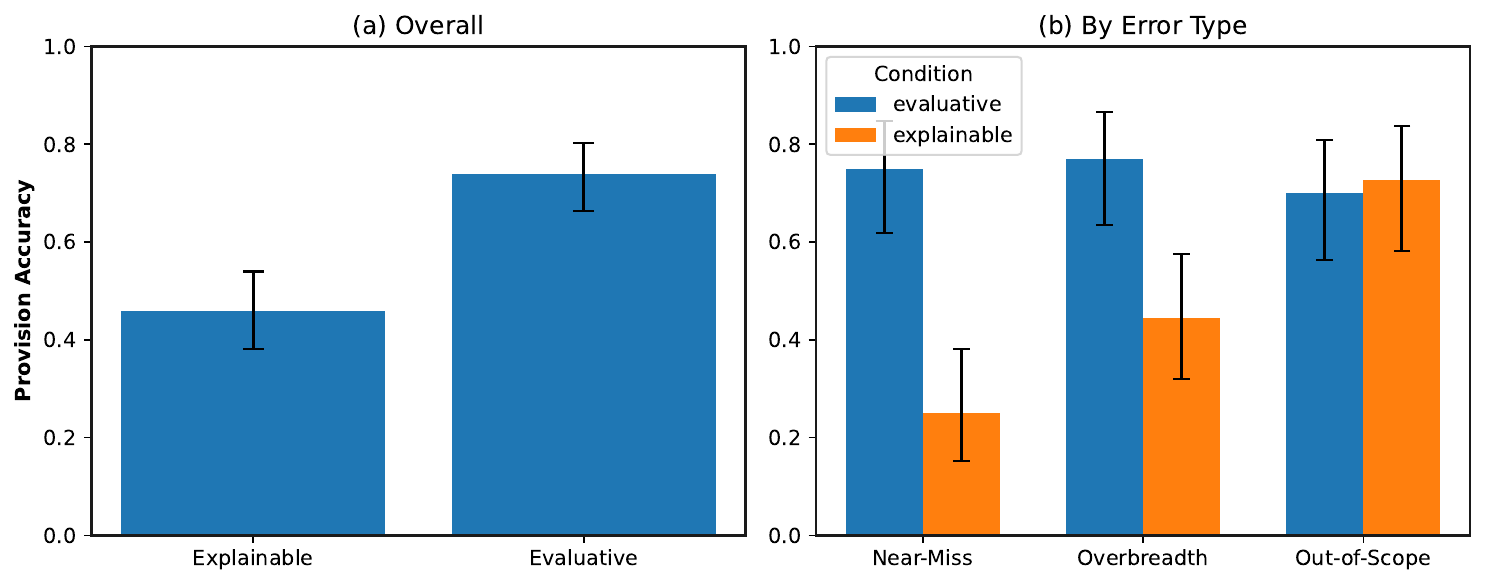}
  \caption{Provision-level accuracy under AI error regimes.
  (a) Overall accuracy on AI-error trials by assistance condition.
  (b) Accuracy by error condition (NM, OB, OS).
  Bars show proportions correct; error bars indicate 95\% Wilson confidence intervals.}
  \Description{Two-panel bar chart of provision-level accuracy on AI error trials comparing Evaluative AI and Conventional XAI (Explainable). Panel (a) shows overall accuracy by condition; panel (b) shows accuracy by error condition. Error bars represent 95\% Wilson confidence intervals.}
  \label{fig:h1_accuracy_overall}
\end{figure}

\textbf{H1: Provision accuracy under AI error}
Evaluative AI substantially improved provision-level accuracy under AI error. Participants in the Evaluative arm were correct on 74.0\% of error trials ($n=150$), compared to 46.0\% in the Conventional XAI arm ($n=150$), a difference of 28 percentage points.
In a logit GLM with heteroskedasticity-robust (HC1) standard errors, including AI error condition and case fixed effects, Evaluative assistance was associated with an approximately ten-fold increase in odds of selecting the correct legal provision relative to Conventional XAI (OR~=~10.96, 95\%~CI [4.35, 27.65], $p<.001$). Results were numerically stable across estimation choices: a covariate-adjusted model including pre-task measures (MAILS, NCS-6, legal knowledge, Propensity to Trust Automation) yielded a comparable effect (OR~=~11.10, 95\%~CI [4.52, 27.20], $p<.001$). 
The Evaluative advantage was present across all error conditions (Fig.~\ref{fig:h1_accuracy_overall}). 
Independent of assistance condition, Out-of-Scope errors were associated with higher baseline accuracy than Near-Miss errors (OR~=~10.34, 95\%~CI [3.77, 28.41], $p<.001$), while Overbreadth errors showed a smaller but still positive shift relative to Near-Miss (OR~=~2.71, 95\%~CI [1.09, 6.72], $p=.032$).
Moderation analyses indicated that the Evaluative advantage was attenuated for Out-of-Scope errors relative to Near-Miss errors (Evaluative$\times$Out-of-Scope: OR~=~0.07, 95\%~CI [0.02, 0.27], $p<.001$), while no reliable attenuation was observed for Overbreadth errors (Evaluative$\times$Overbreadth: OR~=~0.40, 95\%~CI [0.11, 1.46], $p=.165$). 
Case fixed effects were not statistically significant overall; full model tables and balance checks are reported in the Appendix.

\textit{Conclusion.} H1 is supported: when the AI makes an error, Evaluative AI substantially improves provision accuracy, increasing correctness by 28 percentage points and yielding an order-of-magnitude increase in the odds of selecting the correct legal provision relative to Conventional XAI for the reference AI error condition (Near-Miss).

\begin{figure}[h]
  \centering
  \includegraphics[width=1\linewidth]{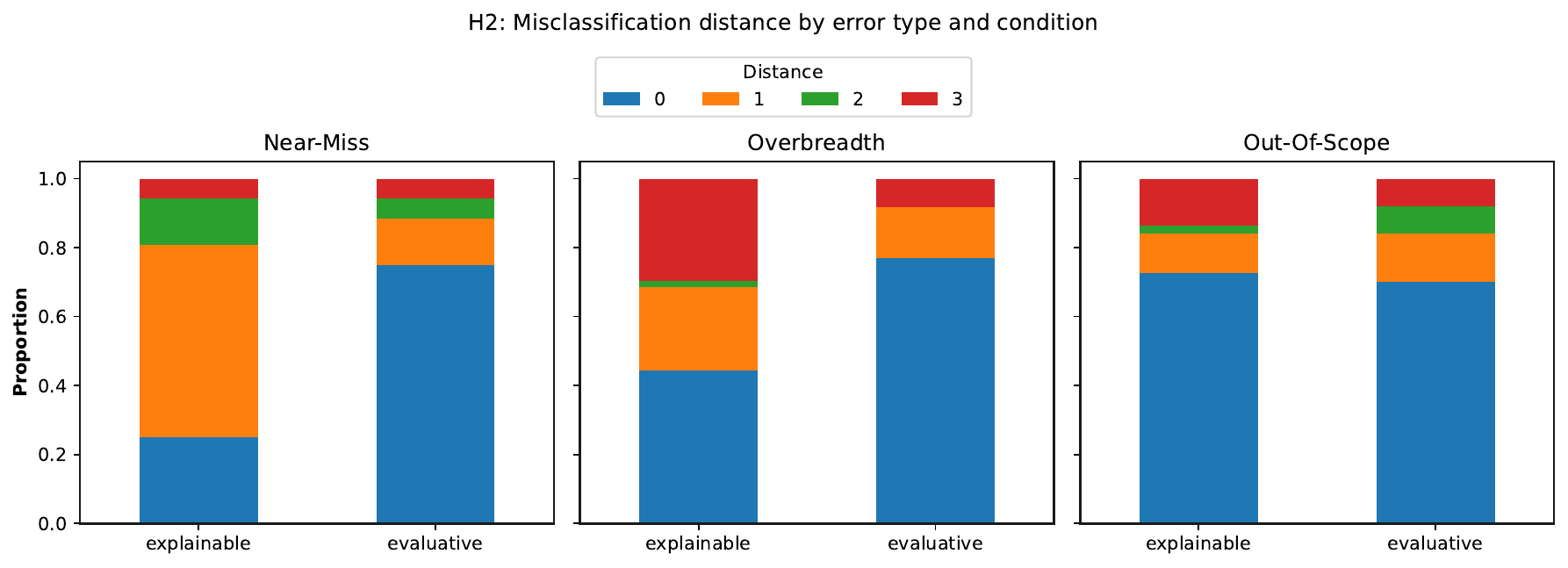}
  \caption{Misclassification distance under AI error by error condition and assistance condition.}
  \Description{Stacked bar charts showing the distribution of misclassification distances (0--3) on error trials, separated by error condition (Near-Miss, Overbreadth, Out-of-Scope) and assistance condition (Conventional XAI vs.\ Evaluative AI). Each bar represents the proportion of trials at each distance level. Across all error conditions, Evaluative AI places relatively more probability mass on smaller distances (0–1) and less on larger distances (2–3) than Conventional XAI. The shift is strongest for Near-Miss errors, comparable for Overbreadth errors, and attenuated for Out-of-Scope errors.}
  \label{fig:h2a_distance_by_condition}
\end{figure}

\textbf{H2: Misclassification distance under AI error}
Evaluative AI reduced misclassification distance relative to Conventional XAI on AI-error trials. In the primary ordered-logit specification (without case fixed effects), the Evaluative coefficient was negative and statistically significant ($\hat\beta=-1.53$, $SE=0.40$, $z=-3.88$, $p<.001$), indicating a shift in probability mass toward smaller (less severe) distances. Interpreted as a proportional-odds effect, this corresponds to $OR=0.22$ with a 95\% confidence interval of $[0.10,\,0.47]$.\footnote{Threshold (cutpoint) parameters are reported in the Appendix but are not substantively interpreted.}
Figure~\ref{fig:h2a_distance_by_condition} visualizes the corresponding distributions: compared to Conventional XAI, Evaluative AI concentrates more probability mass on distances 0--1 and less on distances 2--3.
Error condition moderated the effect. Relative to Near-Miss errors as the reference AI error condition, Out-of-Scope errors were associated with smaller distances overall ($\hat\gamma_{\text{OS}}=-1.34$, $p=.001$). The Evaluative advantage was significantly attenuated for Out-of-Scope errors (Evaluative$\times$Out-of-Scope: $\hat\phi_{\text{Eval}\times\text{OS}}=1.62$, $p=.007$). By contrast, there was no evidence in the interaction tests that Evaluative’s improvement differed between Overbreadth and Near-Miss errors (Evaluative$\times$Overbreadth: $\hat\phi_{\text{Eval}\times\text{OB}}=-0.07$, $p=.898$).
Restricting the sample to Near-Miss and Overbreadth errors (H2b specification), the Evaluative effect remained strong at the Near-Miss baseline ($\hat\beta=-1.61$, $SE=0.40$, $z=-4.02$, $p<.001$; $OR=0.20$, 95\% CI $[0.09,\,0.44]$), while the Evaluative$\times$Overbreadth interaction again was not statistically significant ($\hat\phi_{\text{Eval}\times\text{OB}}=-0.06$, $p=.919$), indicating comparable improvements for Near-Miss and Overbreadth errors.
Results were robust across alternative specifications. Adding covariates (MAILS, NCS-6, legal knowledge, and Propensity to Trust Automation), including case fixed effects as a design-control robustness check, or excluding benign reporters did not alter conclusions. Non-parametric Mann--Whitney tests provided convergent evidence, showing significantly smaller misclassification distances under Evaluative AI overall ($U=8128$, one-tailed $p<.001$), as well as within Near-Miss ($U=734$, $p<.001$) and Overbreadth errors ($U=841.5$, $p=.0003$).

\textit{Conclusion.} H2a is supported: when the AI is wrong, Evaluative AI yields smaller misclassification distances than Conventional XAI. H2b receives partial support: the Evaluative benefit is attenuated for Out-of-Scope errors, while differences between Near-Miss and Overbreadth errors are not detectable in the interaction tests.

\textbf{H3: No-error trials}
\emph{H3a: Accuracy.}
When the AI was correct, provision accuracy was high in both conditions and did not differ reliably between assistance forms. Descriptively, accuracy was 86.7\% under Conventional XAI and 82.0\% under Evaluative AI. In the binomial GLM with HC1 heteroskedasticity-robust standard errors (one observation per participant) and no case fixed effects, the Evaluative vs.\ Conventional XAI contrast was not statistically significant (OR~=~0.65, 95\%~CI [0.35, 1.22], $p=.192$). Results were unchanged when adding case fixed effects as a design-control robustness check.
A TOST equivalence test with margin $\delta=\pm0.05$ did not support equivalence. 
While the upper bound test was satisfied ($p=.011$), the lower bound test failed 
($p=.468$), indicating that equivalence could not be established under the specified margin.
\emph{H3b: Decision time.}
Decision time analyses showed a consistent speed advantage for Conventional XAI. In the OLS model on log-transformed decision times with robust (HC1) standard errors, Evaluative AI was slower than Conventional XAI (coef~=~0.13, $SE=0.07$, $p=.064$), indicating a directionally consistent delay that did not reach conventional significance. Results were substantively unchanged when adding case fixed effects as a design-control robustness check.
The non-parametric Mann--Whitney U test on raw decision times corroborated this directional effect: decisions were significantly faster under Conventional XAI than Evaluative AI ($U=10012$, one-tailed $p=.0498$). Median decision times were 203{,}690\,ms for Conventional XAI and 235{,}769\,ms for Evaluative AI, corresponding to a substantial practical difference.

\textit{Conclusion.}
H3b is supported: when the AI assistance is correct, \xai{} enables faster decisions than \eai{}. H3a remains inconclusive: although the accuracy difference is not statistically significant, equivalence between Evaluative and Conventional XAI is not established under the prespecified margin.

\begin{figure}[t]
  \centering

  \begin{subfigure}[t]{0.48\linewidth}
    \centering
    \includegraphics[width=\linewidth]{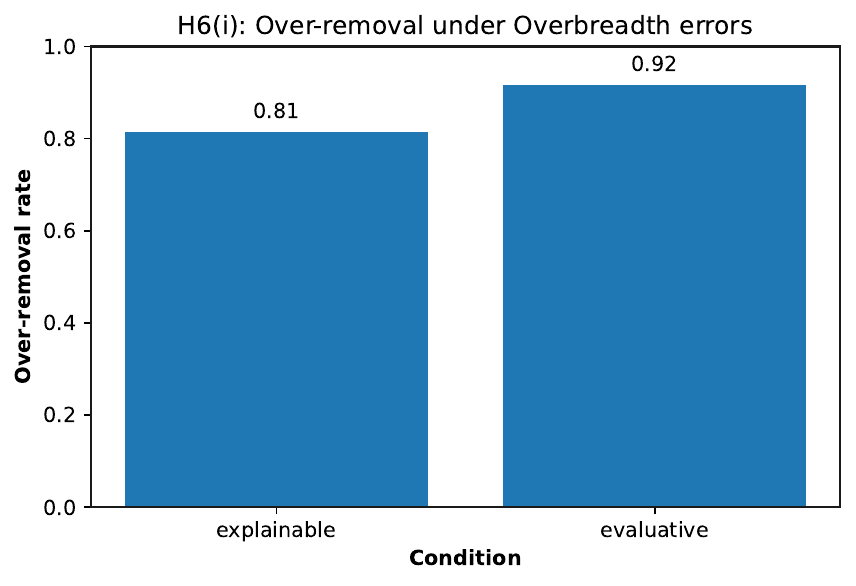}
    \caption{Over-removal under Overbreadth errors.}
    \Description{
    Bar chart comparing over-removal rates under Overbreadth errors between
    Conventional XAI and Evaluative AI. Over-removal denotes selection of severe
    enforcement actions (e.g., removal or suspension). Evaluative AI shows a
    descriptively higher over-removal rate.
    }
    \label{fig:H6_overremoval_overbreadth}
  \end{subfigure}
  \hfill
  \begin{subfigure}[t]{0.48\linewidth}
    \centering
    \includegraphics[width=\linewidth]{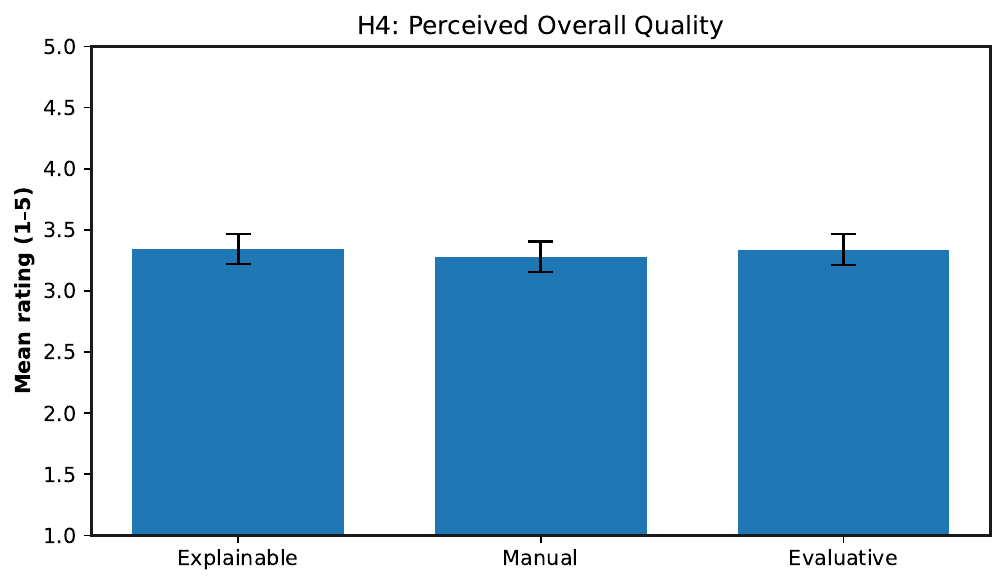}
    \caption{Perceived overall quality of explanations.}
    \Description{
    Bar chart showing mean perceived overall quality ratings (1--5) for \bl{},
    Conventional XAI, and Evaluative AI conditions. Error bars indicate 95\%
    confidence intervals of the mean. Ratings are similar across conditions with
    substantial overlap.
    }
    \label{fig:h4_perceived_quality}
  \end{subfigure}

  \caption{
  Downstream enforcement risk and explanation quality.
  \textbf{(a)} Over-removal rates under Overbreadth errors show no reduction under
  Evaluative AI.
  \textbf{(b)} Perceived overall explanation quality does not differ reliably by
  assistance condition.
  Together, the panels illustrate that gains in decision accuracy do not
  necessarily translate into improved justifications or proportional enforcement.
  }
  \label{fig:h4_h6_combined}
\end{figure}

\textbf{H4: Explanation-of-Reasons (EoR) quality}
We tested whether Evaluative AI improves the quality of users' Explanation-of-Reasons (EoR) submissions
relative to \bl{} reporting and Conventional XAI across four coder-rated dimensions:
element coverage, proportionality reasoning, reasoning depth, and perceived overall quality.
Across all four dimensions, we found no reliable differences between assistance conditions.
In linear regression models on coder-mean ratings with participant-clustered standard errors,
coefficients comparing Evaluative AI to Conventional XAI were small and statistically indistinguishable
from zero (all $|t| < 0.70$, all $p > .48$).
\bl{} reporting likewise did not significantly differ from Conventional XAI on any dimension.
For element coverage, the largest estimated contrast was between Evaluative AI and \bl{} reporting
($\hat\beta = 0.14$, $SE = 0.11$, $t = 1.24$, $p = .216$), but this difference was not statistically significant.
No evidence of differences was observed between Evaluative AI and Conventional XAI
($\hat\beta = 0.007$, $p = .953$).
Figure~\ref{fig:h4_perceived_quality} illustrates the distribution of perceived overall quality ratings.
Distributions for Evaluative AI and Conventional XAI largely overlap across rating levels,
with no systematic shifts consistent with improved explanation quality.

\textit{Conclusion.}
H4 is not supported. Although Evaluative AI improves decision accuracy and error handling (H1--H2), it does not reliably improve coder-rated explanation quality relative to \bl{} reporting or \xai.

\textbf{H5: AI error condition impact and attenuation}
\emph{Accuracy.}
In the paired difference-in-differences logistic model, AI error conditions substantially reduced provision accuracy relative to correct AI trials, and the magnitude of this error penalty varied across AI error conditions.
Relative to Near-Miss as the reference AI error condition, the accuracy decline was significantly smaller for Out-of-Scope errors (error trial$\times$OS $>0$, $p<.001$) and smaller for Overbreadth errors (error trial$\times$OB $>0$, $p=.025$).
Near-Miss errors produced the largest accuracy drops. Both Overbreadth and Out-of-Scope errors were significantly less damaging than Near-Miss errors, with Out-of-Scope errors showing the smallest declines. Differences between Overbreadth and Out-of-Scope were marginal. Wald contrasts confirmed that error penalties differed by AI error condition (OS vs.\ Near-Miss: $p<.001$; OS vs.\ Overbreadth: $p=.054$).
\emph{Attenuation in accuracy.}
Evaluative AI significantly reduced the baseline error penalty relative to Conventional XAI (Evaluative AI$\times$error trial $>0$, $p<.001$), indicating overall attenuation of error-induced accuracy losses.
However, this attenuation was moderated by AI error condition.
The three-way interaction for Out-of-Scope errors was negative and statistically significant (Evaluative AI$\times$error trial$\times$OS $<0$, $p<.001$), indicating that the additional attenuation provided by Evaluative AI was smaller for Out-of-Scope errors than for Near-Miss errors.
No reliable moderation was observed for Overbreadth errors (Evaluative AI$\times$error trial$\times$OB, $p=.232$).
Taken together, Evaluative AI most strongly buffered the accuracy penalty associated with Near-Miss errors, while incremental gains were reduced for AI error conditions that were already less harmful under Conventional XAI.
\emph{EoR quality (coder ratings; first-difference models).}
We next examined whether AI error conditions affected explanation quality differently.
Using participant-level first-difference linear models on continuous coder-mean ratings, error trials were not associated with statistically reliable declines in EoR quality across any dimension.
In particular, Out-of-Scope errors did not produce larger reductions than Near-Miss or Overbreadth errors in \emph{element coverage}, \emph{proportionality reasoning}, \emph{reasoning depth}, or \emph{perceived overall quality} (all $p>.39$).
Accordingly, H5a is not supported for EoR quality.
\emph{Attenuation in EoR quality.}
Evaluative AI did not significantly attenuate error-related changes in explanation quality on any EoR dimension.
Interaction terms between Evaluative AI and Out-of-Scope errors were small and non-significant across all measures (all $p>.39$).
Thus, H5b is not supported for EoR quality.
\emph{Robustness.}
Sensitivity analyses using proportional-odds (ordered logit) models on discretized coder-mean ratings yielded substantively same conclusions, with no sign reversals or emergent significant effects (Appendix).

\emph{Conclusion.}
H5a and H5b are supported for accuracy but not for explanation quality. Evaluative AI selectively attenuates error-induced accuracy losses—most strongly for Near-Miss errors—while providing limited additional benefit for error conditions that are already less harmful under Conventional XAI. Across AI error conditions, however, Evaluative AI does not reliably improve the structure or completeness of user-provided explanations under AI error, indicating that its primary benefit lies in supporting correct decision outcomes rather than downstream explanation quality.

\textbf{H6: DSA-relevant downstream risk outcomes}
\emph{Overbreadth errors (H6(i)).}
The Overbreadth panel comprised 102 error trials from 102 participants (54 Conventional XAI; 48 Evaluative AI). Severe enforcement actions were frequent in both conditions, with higher raw over-removal rates under Evaluative AI (91.7\%) than Conventional XAI (81.5\%).
A binomial GLM with participant-clustered standard errors indicated that the Evaluative AI condition was associated with higher odds of selecting a removal-type action relative to Conventional XAI (log-odds $\beta=0.92$, SE$=0.64$; OR$=2.50$, 95\% CI [0.72, 8.68]), though this difference was not statistically significant ($p=0.149$). Thus, we find no evidence that Evaluative AI reduces disproportionate enforcement under Overbreadth errors.
\emph{Out-of-Scope errors (H6(ii)).}
The Out-of-Scope panel included 94 error trials from 94 participants (50 Evaluative AI; 44 Conventional XAI). Misrouting was near-universal: all Evaluative AI trials (100\%) and nearly all Conventional XAI trials (95.5\%) failed to select the out-of-scope routing option.
Because the outcome exhibited insufficient variation, a regression model could not be meaningfully estimated. We therefore report descriptive statistics only and treat H6(ii) as not identifiable in this dataset.

\emph{Conclusion.}
Overall, H6(i) is not supported: Evaluative AI did not reduce over-removal under Overbreadth errors and showed descriptively higher, though non-significant, rates of severe enforcement. H6(ii) was not identifiable due to near-zero variance in out-of-scope routing behavior, indicating a measurement or interface limitation rather than a treatment effect.
A summary of the hypotheses can be found in Appendix~\ref{summary_H}.

\section{Discussion}
This study examined how different AI assistance architectures shape user reporting behavior in content moderation workflows, focusing on accuracy under AI error, explanation quality, and implications for DSA-relevant reporting objectives. The results reveal a consistent pattern: Evaluative AI improves decision outcomes when the AI is wrong, but does not reliably improve the quality of user-provided explanations. This distinction is consequential for governance-critical settings, \SB{where correct decisions and higher quality user provided justifications do not necessarily coincide.}

\textbf{Reducing overreliance under AI error.}
Under error conditions, Evaluative AI substantially increased provision accuracy (H1) and reduced misclassification distance (H2), indicating less harmful reliance on incorrect model suggestions. These effects are consistent with evaluative assistance functioning as a \emph{deliberation scaffold}: by surfacing competing considerations rather than a single recommended category, the interface makes disagreement explicit and discourages one-shot acceptance of erroneous outputs~\cite{lu2024does, mohammadi2025ai}. Even when users did not fully correct an AI error, evaluative support helped avoid the most severe misclassifications, keeping users' selections closer to legally relevant provisions. For moderation pipelines, this can improve the quality of information entering downstream processes under uncertainty~\cite{wagner_mapping_2024}.

\textbf{Selective attenuation and targeted deployment.}
The paired DiD analyses show that Evaluative AI does not uniformly mitigate all error-related failures. Benefits were strongest for Near-Miss errors, attenuated for Out-of-Scope errors, and not detectably different from Near-Miss for Overbreadth errors. This pattern argues against a one-size-fits-all assistance policy~\cite{thach_trans-centered_2024}. \SB{Instead, the results suggest that evaluative support may be most useful in high-uncertainty or high-risk situations}---for example, when model confidence is low, category entropy is high, or user choices conflict with AI suggestions~\cite{hosamane_i_2025}. Such targeted use could preserve robustness benefits while limiting added time and cognitive load, a key consideration given that evaluative interfaces slowed decisions when the AI was correct (H3b) and full human–AI review is not scalable in high-volume moderation contexts~\cite{lai2022human}.

\textbf{Why explanation quality did not improve.}
Despite accuracy gains, Evaluative AI did not reliably improve coder-rated Explanation-of-Reasons (EoR) quality (H4–H5). This dissociation suggests a mechanism mismatch: evaluative assistance primarily supports \emph{choice revision} (selecting a better provision) rather than \emph{argument construction} (articulating richer justifications). In our design, the reporting interface already required users to engage with statutory elements and evidence fields, which may have elevated baseline explanation quality across conditions and compressed variance. As prior work shows, providing AI explanations or guidance alone does not reliably improve reasoning quality or calibration~\cite{miller2023ai}. Improving EoRs likely requires assistance that directly targets articulation—such as gap-highlighting, prompts for missing elements, or retrieval-backed evidence cues—without generating legal arguments on the user’s behalf.

\textbf{Implications for DSA-oriented design.}
The DSA requires user-friendly reporting mechanisms and sufficiently substantiated notices, but does not specify how interfaces should support users in meeting these standards~\cite{sekwenz_it_2025}. Our findings indicate that Evaluative AI can improve the reliability of user reports under AI error by reducing overreliance and severe misclassification, thereby improving the \emph{quality of information} entering moderation pipelines \SB{at the reporting stage}. However, improved classification accuracy should not be conflated with legally sufficient explanations: Evaluative AI supports \emph{better decisions}, but not necessarily \emph{better justifications}. It is therefore best understood as a performance-enhancing design intervention under uncertainty, not as a mechanism that enforces regulatory compliance~\cite{schmitt_role_2024}.

\textbf{Downstream risks and platform costs.}
Downstream risk proxies (H6) highlight important limits. Under Overbreadth errors, over-removal rates were high in both conditions and descriptively higher under Evaluative AI, with no statistically reliable reduction in disproportionate enforcement. For Out-of-Scope errors, misrouting was near-universal, leaving insufficient variation to identify treatment effects. These results underscore that accuracy gains do not automatically translate into proportionate enforcement and suggest that interface defaults can dominate user choices~\cite{disalvo_social_2022}. More broadly, lowering reporting friction could  increase the volume of actionable notices, raising moderation costs and governance burdens~\cite{cauffman_new_2021}.

\textbf{Limitations and future work.}
This study relied on a custom reporting interface with a curated set of legal provisions, which may have reduced real-world ambiguity and increased baseline performance relative to deployment settings. The structured explanation fields, while useful for measurement, may scaffold reasoning in ways that differ from deployed platform environments. Moreover, the experiment captures short-term behavior and cannot assess learning, strategic adaptation, or adversarial misuse over time. Future work should evaluate evaluative assistance in platform-integrated settings, link user reports to moderator outcomes, and explore adaptive or constrained writing support that better aligns gains in decision accuracy with meaningful improvements in explanation quality and downstream governance outcomes.

\section{Conclusion}

This study demonstrates that Evaluative AI improves user accuracy under AI error in illegal content reporting by reducing harmful overreliance relative to conventional XAI. Whereas conventional XAI interfaces encourage one-shot acceptance of incorrect suggestions and increase susceptibility to biases favorable to machine outputs, Evaluative AI supports more cautious and contestable decision-making under uncertainty. At the same time, we find no reliable improvements in the quality of user-provided explanations, indicating that gains in decision accuracy do not automatically translate into more substantiated or higher-quality reasoning.

These findings underscore that reporting interfaces are an important design space under the Digital Services Act and can improve performance under the right conditions. Evaluative AI is best understood as a deliberation scaffold that improves decision outcomes in error-prone settings, rather than as a mechanism that guarantees legally sufficient explanations or independently satisfies regulatory requirements. Important trade-offs remain: evaluative assistance increases deliberation time and may raise the volume of actionable reports, with implications for platform costs and downstream governance burden.

Future work should examine longitudinal effects on trust, learning, and strategic behavior; test generalizability across jurisdictions and platform contexts; and explore complementary support modalities—such as adaptive or constrained assistance—that directly target explanation articulation. Advancing such designs will be critical for balancing accuracy, user agency, and regulatory accountability in human–AI collaboration.

\bibliographystyle{ACM-Reference-Format}
\bibliography{CHI_Lit}

\section{Generative AI Usage Statement}
In line with the ACM policy and FAccT guidelines, the authors used ChatGPT to refine the phrasing of certain sentences. They were also employed in drafting portions of the programming work, such as elements of the experimental interface, data analysis tools, and figure generation scripts (see Appendix). We did not use generative AI to generate substantive content, arguments, results, analyses, or interpretations. All suggested edits were reviewed, selectively adopted, and verified by the authors, who remain fully responsible for the final manuscript.
\appendix

\section{Summary of Hypothesis}
\label{summary_H}

\begin{table}[H]
  \centering
  \footnotesize
  \caption{Hypotheses overview.}
  \label{tab:hypotheses}
  \begin{tabular}{p{2.3cm} p{11cm}}
    \toprule
    \textbf{Hypothesis} & \textbf{Statement} \\
    \midrule
    H1 & Under AI error, Evaluative AI yields higher provision-level accuracy than conventional XAI (less overreliance on incorrect outputs). \\
    \midrule
    H2a & Evaluative AI reduces misclassification distance relative to XAI. \\
    H2b & The H2a reduction varies by error type; the Evaluative effect is attenuated for Out-of-Scope errors relative to Near-Miss/Overbreadth. \\
    \midrule
    H3a & When the AI is correct, Evaluative and XAI have comparable accuracy. \\
    H3b & When the AI is correct, XAI decisions are faster than Evaluative. \\
    \midrule
    H4 & Evaluative AI elicits higher-quality explanations than Manual and XAI. \\
    \midrule
    H5a & AI error types differ in the magnitude of their impact on reporting performance. \\
    H5b & Evaluative AI differentially attenuates error-type–specific performance drops relative to XAI. \\

    \midrule
    H6 & Versus XAI, Evaluative better aligns with DSA goals: (i) lower over-removal under Overbreadth; (ii) lower misrouting under OS. \\
    \bottomrule
  \end{tabular}
\end{table}

\begin{table*}[H]
\centering
\footnotesize
\caption{Summary of hypotheses, tests, findings, and conclusions.}
\label{tab:hypotheses_summary}
\begin{tabular}{p{1.2cm} p{6.2cm} p{6.0cm}}
\toprule
\textbf{Hyp.} & \textbf{Tested} & \textbf{Findings and conclusion} \\
\midrule
H1 &
Provision accuracy on AI-error trials (Evaluative vs.\ Conventional XAI), with error subtype and case fixed effects. &
\textbf{Supported.} Evaluative improved accuracy by 28pp (74\% vs.\ 46\%). In logit GLM (HC1) with error-subtype and case fixed effects, Evaluative increased odds of correct provision selection (OR~=~10.96, 95\% CI [4.35, 27.65], $p<.001$). \\
\midrule
H2a &
Misclassification distance $D\in\{0,1,2,3\}$ on AI-error trials (ordered logit; no case FE primary). &
\textbf{Supported.} Evaluative reduced misclassification distance (ordered logit $\hat\beta=-1.53$, $SE=0.40$, $p<.001$; OR~=~0.22, 95\% CI [0.10, 0.47]), shifting mass toward smaller distances (0--1). \\
H2b &
Condition $\times$ error subtype moderation in ordered logit; plus focused NM vs.\ OB model. &
\textbf{Partially supported.} Evaluative benefit was \emph{attenuated} for Out-of-Scope errors (Eval$\times$OOS $\hat\phi=1.62$, $p=.007$). No detectable difference between Overbreadth and Near-Miss attenuation (Eval$\times$OB $p=.898$; focused NM/OB interaction $p=.919$). \\
\midrule
H3a &
Provision accuracy on no-error trials (AI correct): GLM (HC1) + TOST equivalence test. &
\textbf{Inconclusive.} Accuracy did not differ significantly (OR~=~0.65, 95\% CI [0.35, 1.22], $p=.192$), but equivalence was not established under $\delta=\pm0.05$ (upper bound passed; lower bound failed). \\
H3b &
Decision time on no-error trials: OLS on log DT (HC1) + Mann--Whitney U on raw DT. &
\textbf{Supported.} Conventional XAI was faster. OLS showed a directional delay for Evaluative (coef~=~0.13, $p=.064$) and MWU corroborated faster XAI decisions ($U=10012$, one-tailed $p=.0498$; medians 203{,}690ms vs.\ 235{,}769ms). \\
\midrule
H4 &
Coder-rated EoR quality (4 dimensions): linear models with participant-clustered SEs. &
\textbf{Not supported.} No reliable differences between conditions across any dimension (all $|t|<0.70$, all $p>.48$). \\
\midrule
H5a &
Paired DiD (AI arms): error penalty by error type for accuracy and EoR quality. &
\textbf{Supported for accuracy; not supported for EoR.}
Error trials reduced accuracy across types, with penalties differing by error type (Near-Miss largest drop; Overbreadth intermediate; Out-of-Scope least damaging; Wald contrasts $p<.05$).
For EoR quality, first-difference models showed no statistically reliable error-type effects on any dimension (all $p>.39$). \\
\midrule
H5b &
Paired DiD three-way: Condition $\times$ post $\times$ ErrorType attenuation. &
\textbf{Supported for accuracy in a targeted manner; not supported for EoR.}
Evaluative reduced the overall error penalty (Condition$\times$post, $p<.001$), but attenuation was \emph{smaller} for Out-of-Scope than for Near-Miss (Eval$\times$post$\times$OOS $<0$, $p<.001$) and not reliably different for Overbreadth ($p=.232$). No reliable attenuation was observed for EoR dimensions (all $p>.39$). \\
\midrule
H6(i) &
Over-removal under Overbreadth errors: GLM with participant-clustered SEs. &
\textbf{Not supported.} Over-removal was high in both conditions and descriptively higher under Evaluative (91.7\% vs.\ 81.5\%); GLM OR~=~2.50 (95\% CI [0.72, 8.68], $p=.149$). \\
H6(ii) &
Misrouting under Out-of-Scope errors. &
\textbf{Not identifiable.} Misrouting was near-universal (Evaluative 100\%, XAI 95.5\%), leaving insufficient variance for meaningful regression; descriptive reporting only. \\
\bottomrule
\end{tabular}
\end{table*}

\section{H3: Decision time on no-error trials by assistance condition}

\begin{figure}[H]
  \centering
  \includegraphics[width=0.6\linewidth]{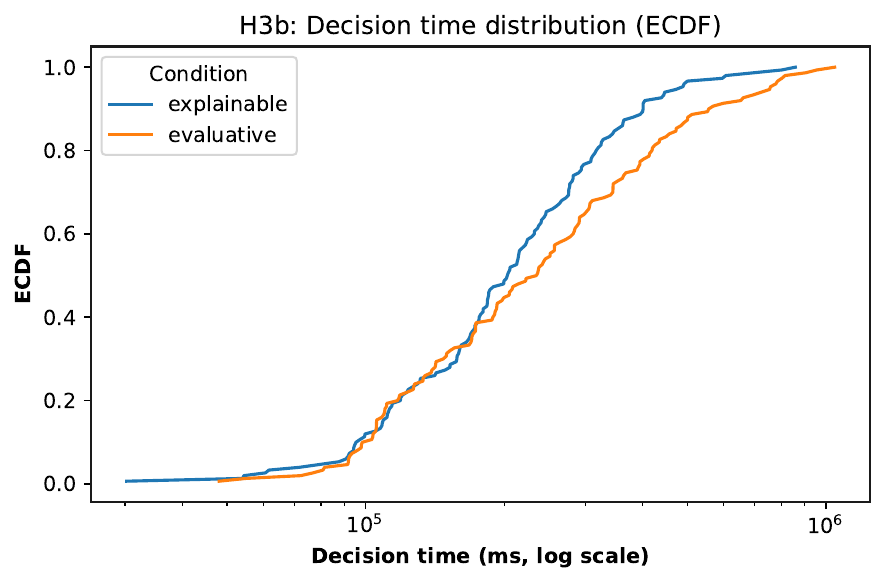}
  \caption{Decision time on no-error trials by assistance condition.}
  \Description{Empirical cumulative distribution functions (ECDFs) of decision times on no-error trials, plotted on a log-scaled x-axis. At any given time threshold, a larger proportion of Conventional XAI decisions are completed relative to Evaluative AI, indicating a robust distribution-wide speed advantage when the AI output is correct.}
  \label{fig:h3_decision_time}
\end{figure}

\newpage

\section{Error Types}
\begin{table*}[h]
  \caption{Error types (Near-miss, Overbreadth, Out-of-scope) for selected legal content categories.}
  \label{tab:error_examples}
 \scriptsize
 \setlength{\tabcolsep}{2.2pt}
  \centering
  \begin{tabular}{
    >{\raggedright\arraybackslash}p{0.22\linewidth}
    >{\raggedright\arraybackslash}p{0.22\linewidth}
    >{\raggedright\arraybackslash}p{0.22\linewidth}
    >{\raggedright\arraybackslash}p{0.22\linewidth}}
    \toprule
    \textbf{Provision} & \textbf{Near-miss} & \textbf{Overbreadth} & \textbf{Out-of-scope} \\
    \midrule
    Incitement to crime (\S~111) &
    Prohibited symbols (\S~86a), Breach of peace (\S~126) &
    Incitement to hatred (\S~130), Criminal organizations (\S~129) &
    Privacy images (\S~201a), Defamation (\S~186--188) \\
    \midrule
    Incitement to hatred (\S~130) &
    Incitement to crime (\S~111), Encouraging violence (\S~91) &
    Insult (\S~185--188), Religious defamation (\S~166) &
    Privacy images (\S~201a), Data forgery (\S~269) \\
    \midrule
    Use of prohibited symbols (\S~86a) &
    Propaganda (\S~86), Incitement to crime (\S~111) &
    Incitement to hatred (\S~130), Encouraging violence (\S~91) &
    Privacy images (\S~201a), Child pornography (\S~184b) \\
    \midrule
    ``Something else'' (placeholder) &
    Incitement to crime (\S~111), Data forgery (\S~269) &
    Incitement to hatred (\S~130), Privacy images (\S~201a) &
    Prohibited symbols (\S~86a), Child pornography (\S~184b) \\
    \bottomrule
  \end{tabular}
\end{table*}

\section{Pre-Study Questionnaire}
\label{questionnaire}
Please answer each question to the best of your ability. This helps us understand your background and experience.

\subsection{Need for Cognition (5-point Likert)}
\noindent\textbf{Answer options:} 1 = Strongly Disagree; 2 = Disagree; 3 = Neutral; 4 = Agree; 5 = Strongly Agree.

\begin{enumerate}
  \item I would prefer complex to simple problems. \emph{(Need for Cognition — Question 1)}
  \item I like to have the responsibility of handling a situation that requires a lot of thinking. \emph{(Need for Cognition — Question 2)}
  \item Thinking is not my idea of fun. \emph{(Need for Cognition — Question 3)}
  \item This is an attention check question, please select "Strongly Agree". \emph{(Attention Check — Question 4)}
  \item I would rather do something that requires little thought than something that is sure to challenge my thinking abilities. \emph{(Need for Cognition — Question 5)}
   \item I really enjoy a task that involves coming up with new solutions to problems. \emph{(Need for Cognition — Question 6)}
  \item I would prefer a task that is intellectual, difficult, and important to one that is somewhat important but does not require much thought. \emph{(Need for Cognition — Question 7)}
\end{enumerate}

\subsection{Use \& Apply AI (0–10 scale)}
\noindent\textbf{Select one:} 0, 1, 2, 3, 4, 5, 6, 7, 8, 9, 10. \quad (0 = Not at all; 10 = Extremely well.)

\begin{enumerate}
  \item I can operate AI applications in everyday life. \emph{(Use \& Apply AI — Question 1)}
  \item I can use AI applications to make my everyday life easier. \emph{(Use \& Apply AI — Question 2)}
  \item I can use artificial intelligence meaningfully to achieve my everyday goals. \emph{(Use \& Apply AI — Question 3)}
   \item To ensure data quality, please rate this item as 7 out of 10. \emph{(Attention Check — Question 4)}
  \item In everyday life, I can interact with AI in a way that makes my tasks easier. \emph{(Use \& Apply AI — Question 5)}
  \item In everyday life, I can work together gainfully with an artificial intelligence. \emph{(Use \& Apply AI — Question 6)}
  \item I can communicate gainfully with artificial intelligence in everyday life. \emph{(Use \& Apply AI — Question 7)}
\end{enumerate}

\subsection{Know \& Understand AI (0–10 scale)}
\noindent\textbf{Select one:} 0, 1, 2, 3, 4, 5, 6, 7, 8, 9, 10. \quad (0 = Not at all; 10 = Extremely well.)

\begin{enumerate}
  \item I know the most important concepts of the topic ``artificial intelligence''. \emph{(Know \& Understand AI — Question 1)}
  \item I know definitions of artificial intelligence. \emph{(Know \& Understand AI — Question 2)}
  \item I can assess what the limitations and opportunities of using an AI are. \emph{(Know \& Understand AI — Question 3)}
  \item I can assess what advantages and disadvantages the use of an artificial intelligence entails. \emph{(Know \& Understand AI — Question 4)}
  \item I can think of new uses for AI. \emph{(Know \& Understand AI — Question 5)}
  \item I can imagine possible future uses of AI. \emph{(Know \& Understand AI — Question 6)}
\end{enumerate}

\subsection{Detect AI (0–10 scale)}
\noindent\textbf{Select one:} 0, 1, 2, 3, 4, 5, 6, 7, 8, 9, 10. \quad (0 = Not at all; 10 = Extremely well.)

\begin{enumerate}
  \item I can tell if I am dealing with an application based on artificial intelligence. \emph{(Detect AI — Question 1)}
  \item I can distinguish devices that use AI from devices that do not. \emph{(Detect AI — Question 2)}
  \item I can distinguish if I interact with an AI or a ``real human''. \emph{(Detect AI — Question 3)}
\end{enumerate}

\subsection{AI Ethics (0–10 scale)}
\noindent\textbf{Select one:} 0, 1, 2, 3, 4, 5, 6, 7, 8, 9, 10. \quad (0 = Not at all; 10 = Extremely well.)

\begin{enumerate}
  \item I can weigh the consequences of using AI for society. \emph{(AI Ethics — Question 1)}
  \item I can incorporate ethical considerations when deciding whether to use data provided by an AI. \emph{(AI Ethics — Question 2)}
  \item I can analyze AI-based applications for their ethical implications. \emph{(AI Ethics — Question 3)}
\end{enumerate}

\subsection{AI Self-Efficacy — Problem Solving (0–10 scale)}
\noindent\textbf{Select one:} 0, 1, 2, 3, 4, 5, 6, 7, 8, 9, 10. \quad (0 = Not at all; 10 = Extremely well.)

\begin{enumerate}
  \item I can rely on my skills in difficult situations when using AI. \emph{(AI Self-Efficacy — Problem Solving — Question 1)}
  \item I can handle most problems in dealing with artificial intelligence well on my own. \emph{(AI Self-Efficacy — Problem Solving — Question 2)}
  \item I can also usually solve strenuous and complicated tasks when working with artificial intelligence well. \emph{(AI Self-Efficacy — Problem Solving — Question 3)}
\end{enumerate}

\subsection{AI Self-Efficacy — Learning (0–10 scale)}
\noindent\textbf{Select one:} 0, 1, 2, 3, 4, 5, 6, 7, 8, 9, 10. \quad (0 = Not at all; 10 = Extremely well.)

\begin{enumerate}
  \item I can keep up with the latest innovations in AI applications. \emph{(AI Self-Efficacy — Learning — Question 1)}
  \item Despite the rapid changes in the field of artificial intelligence, I can always keep up to date. \emph{(AI Self-Efficacy — Learning — Question 2)}
  \item Although there are often new AI applications, I manage to always be ``up-to-date''. \emph{(AI Self-Efficacy — Learning — Question 3)}
\end{enumerate}

\subsection{AI Self-Competency — Persuasion Literacy (0–10 scale)}
\noindent\textbf{Select one:} 0, 1, 2, 3, 4, 5, 6, 7, 8, 9, 10. \quad (0 = Not at all; 10 = Extremely well.)

\begin{enumerate}
  \item I don't let AI influence me in my everyday decisions. \emph{(AI Self-Competency — Persuasion Literacy — Question 1)}
  \item I can prevent an AI from influencing me in my everyday decisions. \emph{(AI Self-Competency — Persuasion Literacy — Question 2)}
  \item I realise if artificial intelligence is influencing me in my everyday decisions. \emph{(AI Self-Competency — Persuasion Literacy — Question 3)}
\end{enumerate}

\subsection{AI Self-Competency — Emotion Regulation (0–10 scale)}
\noindent\textbf{Select one:} 0, 1, 2, 3, 4, 5, 6, 7, 8, 9, 10. \quad (0 = Not at all; 10 = Extremely well.)

\begin{enumerate}
  \item I keep control over feelings like frustration and anxiety while doing everyday things with AI. \emph{(AI Self-Competency — Emotion Regulation — Question 1)}
  \item I can handle it when everyday interactions with AI frustrate or frighten me. \emph{(AI Self-Competency — Emotion Regulation — Question 2)}
  \item I can control my euphoria that arises when I use artificial intelligence for everyday purposes. \emph{(AI Self-Competency — Emotion Regulation — Question 3)}
\end{enumerate}

\subsection{Trust in Automation (5-point Likert)}
\label{TiA}
\noindent\textbf{Answer options:} 1 = Strongly Disagree; 2 = Disagree; 3 = Neutral; 4 = Agree; 5 = Strongly Agree.

\begin{enumerate}
  \item One should be careful with unfamiliar automated systems. \emph{(Trust in Automation — Question 1)}
  \item I rather trust a system than I mistrust it. \emph{(Trust in Automation — Question 2)}
   \item For quality control purposes, please select "Disagree" for this statement. \emph{(Attention Check — Question 3)}
  \item Automated systems generally work well. \emph{(Trust in Automation — Question 4)}
\end{enumerate}

\subsection{Legal Knowledge (5-point Likert)}
\label{Legal_Knowledge}
\noindent\textbf{Answer options:} 1 = Strongly Disagree; 2 = Disagree; 3 = Neutral; 4 = Agree; 5 = Strongly Agree.

\begin{enumerate}
  \item I am familiar with basic legal restrictions on online content in my country. \emph{(Legal Knowledge — Question 1)}
  \item I can identify when a post likely violates the law. \emph{(Legal Knowledge — Question 2)}
  \item I know where to look up legal rules for online speech. \emph{(Legal Knowledge — Question 3)}
\end{enumerate}

\section{Post-Study Questionnaire}
\label{post_questionnaire}

\subsection{Usefulness and Effectiveness (7-point Likert)}
\noindent\textbf{Answer options:} 1 = Strongly Disagree; 2 = Disagree; 3 = Somewhat Disagree; 4 = Neutral; 5 = Somewhat Agree; 6 = Agree; 7 = Strongly Agree.

\begin{enumerate}
  \item Using the AI assistance enabled me to accomplish the reporting task more quickly. \emph{(Usefulness — Effectiveness — Question 1)}
  \item Using the AI assistance improved the quality of my reporting decisions. \emph{(Usefulness — Effectiveness — Question 2)}
  \item Using the AI assistance increased my productivity in completing the task. \emph{(Usefulness — Productivity — Question 3)}
  \item Using the AI assistance enhanced my effectiveness in the reporting task. \emph{(Usefulness — Task Effectiveness — Question 4)}
  \item Using the AI assistance made the reporting task easier to complete. \emph{(Usefulness — Ease — Question 5)}
  \item I found the AI assistance useful for the reporting task. \emph{(Usefulness — Overall — Question 6)}
\end{enumerate}

\subsection{Workload (7-point Likert)}
\noindent\textbf{Answer options:} 1 = Very Low; 2 = Low; 3 = Somewhat Low; 4 = Neutral; 5 = Somewhat High; 6 = High; 7 = Very High.

\begin{enumerate}
  \item How mentally demanding was the reporting task with AI assistance? \emph{(Workload — Mental Demand — Question 1)}
  \item How much effort did you have to expend to accomplish the task with AI assistance? \emph{(Workload — Effort — Question 2)}
  \item How frustrated did you feel during the task with AI assistance? \emph{(Workload — Frustration — Question 3)}
\end{enumerate}

\subsection{Trust, Control, and Responsibility (7-point Likert)}
\noindent\textbf{Answer options:} 1 = Strongly Disagree; … 7 = Strongly Agree.

\begin{enumerate}
  \item I felt in control of my reporting decisions when using the AI assistance. \emph{(Trust/Control — Question 1)}
  \item I felt responsible for the outcomes of my reporting decisions when using AI assistance. \emph{(Responsibility — Question 2)}
  \item I felt like I was just following the AI suggestions rather than making my own decisions. \emph{(Reliance — Question 3)}
  \item I could easily override or ignore the AI's suggestions when I wanted to. \emph{(Trust/Control — Override — Question 4)}
\end{enumerate}

\subsection{Demographics and Final Comments}
\begin{enumerate}
  \item Please indicate your age. \emph{(Demographics — Question 1)}
  \item Please indicate your gender. \emph{(Demographics — Question 2)}
  \item Any final comments about your experience? \emph{(Open Feedback — Question 1)}
\end{enumerate}

\begin{enumerate}
  \item Age. \emph{(Demographic Data Collection — Question 1)}
  \item Gender. \emph{(Demographic Data Collection — Question 2)}
  \item Any final comments? \emph{(Optional Feedback — Question 3)}
\end{enumerate}

\section{Descriptives for Participants, Error Types, and Legal Conditions}
\begin{table}[H]
\centering
\footnotesize
\caption{Descriptives for participants, error types (malicious trials), and case law stimuli.}
\label{tab:descriptives}
\begin{tabular}{lccc}
\toprule
 & \textbf{Evaluative} & \textbf{Explainable} & \textbf{Manual} \\
\midrule
Participants (valid) & 150 & 150 & 150 \\
\midrule
\multicolumn{4}{l}{\textit{Error type counts (malicious trials; AI arms only)}} \\
\quad Near-Miss (NM) & 52 & 52 & -- \\
\quad Overbreadth (OB) & 48 & 54 & -- \\
\quad Out-of-Scope (OS) & 50 & 44 & -- \\
\midrule
\multicolumn{4}{l}{\textit{Demographics (valid $N{=}450$)}} \\
\quad Gender (m / f / nb) & \multicolumn{3}{c}{293 / 152 / 5} \\
\quad Age & \multicolumn{3}{c}{$M{=}31.32$, $SD{=}8.7$, range 18--71} \\
\midrule
\multicolumn{4}{l}{\textit{Case law exposure (number of participants; AI arms)}} \\
\quad \S~86a (symbols) & \multicolumn{3}{c}{92} \\
\quad \S~111 (incitement to crime) & \multicolumn{3}{c}{76} \\
\quad \S~130 (incitement to hatred) & \multicolumn{3}{c}{71} \\
\quad Other (e.g., \S~30a BtMG) & \multicolumn{3}{c}{61} \\
\bottomrule
\end{tabular}
\end{table}

\newpage
\section{Codebook for Explaination of the Reason Quality and Related Measures}
\label{Codebook}
\begin{enumerate}
    \item \textbf{Element Coverage}  
    \begin{itemize}
        \item \emph{Definition:} Extent to which participants include all required legal elements in their explanation.  
        \item \emph{Item:} ``The explanation covered all legally relevant elements required for this category.''  
        \item \emph{Scale:}  
        \begin{enumerate}
            \item[1] None of the elements covered  
            \item[2] Few elements covered (1–2)  
            \item[3] About half covered  
            \item[4] Most covered (more than half)  
            \item[5] All covered comprehensively  
        \end{enumerate}
    \end{itemize}

    \item \textbf{Proportionality Reasoning}  
    \begin{itemize}
        \item \emph{Definition:} Depth, proportionality, and correctness of reasoning.  
        \item \emph{Item:} ``The explanation was clear, accurate, and legally proportional.''  
        \item \emph{Scale:}  
        \begin{enumerate}
            \item[1] Very poor (unclear, irrelevant, incorrect)  
            \item[2] Weak (minimal detail, vague, partially incorrect)  
            \item[3] Moderate (some detail, limited proportionality)  
            \item[4] Good (clear, mostly correct, proportionate)  
            \item[5] Excellent (comprehensive, precise, legally sound)  
        \end{enumerate}
    \end{itemize}

    \item \textbf{Reasoning Depth}  
    \begin{itemize}
        \item \emph{Definition:} Extent of elaboration and integration of legal/contextual reasoning.  
        \item \emph{Item:} ``The explanation demonstrates deep reasoning and elaboration.''  
        \item \emph{Scale:}  
        \begin{enumerate}
            \item[1] Very shallow (minimal or no reasoning, single-word/phrase)  
            \item[2] Shallow (basic, limited detail, surface-level)  
            \item[3] Moderate (some elaboration, partial justification)  
            \item[4] Deep (multiple relevant points, good detail)  
            \item[5] Very deep (comprehensive, nuanced arguments)  
        \end{enumerate}
    \end{itemize}

    \item \textbf{Perceived Overall Quality}  
    \begin{itemize}
        \item \emph{Definition:} Rater’s holistic judgment of clarity, persuasiveness, and adequacy.  
        \item \emph{Item:} ``Overall, how good was the explanation of the reason?''  
        \item \emph{Scale:}  
        \begin{enumerate}
            \item[1] Very poor (unclear, incoherent)  
            \item[2] Poor (vague, with errors)  
            \item[3] Moderate (adequate but incomplete)  
            \item[4] Good (clear, coherent, minor gaps)  
            \item[5] Very good (comprehensive, persuasive, legally sound)  
        \end{enumerate}
    \end{itemize}

    \item \textbf{Moderation Decision}  
    \begin{itemize}
        \item \emph{Definition:} The action recommended by the participant.  
        \item \emph{Options:} Remove, Escalate, Warn, Suspend, or No action.  
    \end{itemize}

    \item \textbf{Notes (Optional)}  
    \begin{itemize}
        \item \emph{Definition:} Free-text space for coders to note ambiguities, uncertainties, or context.  
        \item \emph{Examples:} ``Participant confused §111 with §130 but reasoning was detailed.''  
    \end{itemize}
\end{enumerate}

\section{Prompting Template for AI generated suggestions for condition 2}

\begin{itemize}
  \item \textbf{Be explicit about the model’s role and instructions.}  A well-crafted prompt should specify the task, the context and the criteria the model must follow \cite{palla2025policyaspromptrethinkingcontentmoderation, elaraby-etal-2024-persuasiveness}.  The literature recommends a three-part prompt with a system message, task description and formatting examples \cite{elaraby-etal-2024-persuasiveness}.
  \item \textbf{Provide structured and concise instructions.}  Structured prompts using bullet points and clear sections improve moderation accuracy; verbose prompts reduce it \cite{palla2025policyaspromptrethinkingcontentmoderation}.  The template therefore uses numbered steps and bullet points.
  \item \textbf{Include examples to guide output.}  In-context examples of compliant and non-compliant content help the model contextualise the task\cite{palla2025policyaspromptrethinkingcontentmoderation} and avoid misclassification.  The template asks the experimenter to insert examples showing both reportable and non-reportable content.
  \item \textbf{Generate arguments for and against the decision.}  ArgLLMs provide arguments for and against particular outputs, yielding explainable and contestable decisions \cite{freedman2025argumentativelargelanguagemodels}.  The template requires each agent to provide at least one argument supporting and one opposing the reporting decision, including references to the relevant ToS and legal rules.
  \item \textbf{Use multiple LLM agents or perspectives.}  Multi-agent reasoning frameworks (e.g., SocraSynth) use two LLM agents representing opposing viewpoints, moderated by a human \cite{chang2024socrasynthmultillmreasoningconditional}.  Including in our paper both a platform-leaning and a regulator-leaning agent helps surface biases and promotes value pluralism.
  \item \textbf{Avoid randomness and ensure reproducibility.}  To reduce variability in responses, Elaraby et al set the generation temperature to 0 \cite{elaraby-etal-2024-persuasiveness}.  The template instructs the experimenter to set a low temperature or deterministic sampling.
  \item \textbf{Make the output accessible and user-friendly.}  The Digital Services~Act requires that terms and conditions be presented in clear, plain, intelligible language and reporting mechanisms user-friendly and easy to access.  The template instructs the LLM to use simple language and avoid legal jargon where possible, while still citing the relevant rule.
  \item \textbf{Keep response lengths consistent across scenarios.}  To avoid length-based bias and to ensure comparability, responses must have roughly the same number of words \cite{palla2025policyaspromptrethinkingcontentmoderation}.  The template specifies word-count targets for each section.
\end{itemize}

\section{Selection of and creation of reporting scenarios and  legal cases relevant to online content reporting in Germany}

\begin{table*}[h]
  \caption{Examples of illegal content categories, sources, case law, and posts.}
     \label{tab:posts}
  \scriptsize
  \setlength{\tabcolsep}{2.5pt}
  \centering
  \setlength{\tabcolsep}{4pt}
  \begin{tabular}{@{}%
    >{\raggedright\arraybackslash}p{0.06\textwidth}%
    >{\raggedright\arraybackslash}p{0.15\textwidth}%
    >{\raggedright\arraybackslash}p{0.09\textwidth}%
    >{\raggedright\arraybackslash}p{0.12\textwidth}%
    >{\raggedright\arraybackslash}p{0.45\textwidth}@{}}
    \toprule
    \textbf{Num.} & \textbf{Category} & \textbf{Hard/Easy} & \textbf{Case law} & \textbf{Post on SocialNet} \\
    \midrule
    L\textunderscore 6 &
    Using symbols of unconstitutional and terrorist organizations &
    Easy &
    \S~86a German Criminal Code &
    \emph{Finally got this retro banner up in my room. Looks awesome! [photo shows a flag with a prohibited emblem].}
    \par \emph{New profile pic! Proud of my heritage — some symbols carry power. [prohibited symbol image in avatar].} \\
    \midrule
    L\textunderscore 10 &
    Public incitement to crime &
    Hard &
    \S~111 German Criminal Code &
    \emph{Share this: at dawn we smash their labs and shut them down for good.}
    \par \emph{They won’t listen unless we burn down their offices tomorrow. Who’s in?} \\
    \midrule
    L\textunderscore 15 &
    Incitement to hatred &
    Hard &
    \S~130 German Criminal Code &
    \emph{Foreigners are parasites ruining everything — they should be driven out!}
    \par \emph{Refugees don’t belong here — they’re an infestation. Get them out.} \\
    \midrule
    L\textunderscore 27 &
    Something else &
    Easy &
    (e.g., \S~30a Abs.\ 1 BtMG) &
    \emph{Selling unlicensed prescription antibiotics — message me for prices and pickup.}
    \par \emph{Got extra opioid pain meds, no prescription needed. DM to buy discreetly.} \\
    \bottomrule
  \end{tabular}
\end{table*}

\subsection{Screenshot of Facebook's Reporting UI for Illegal Content}

\begin{figure}[H]
  \centering
  \includegraphics[width=0.63\textwidth]{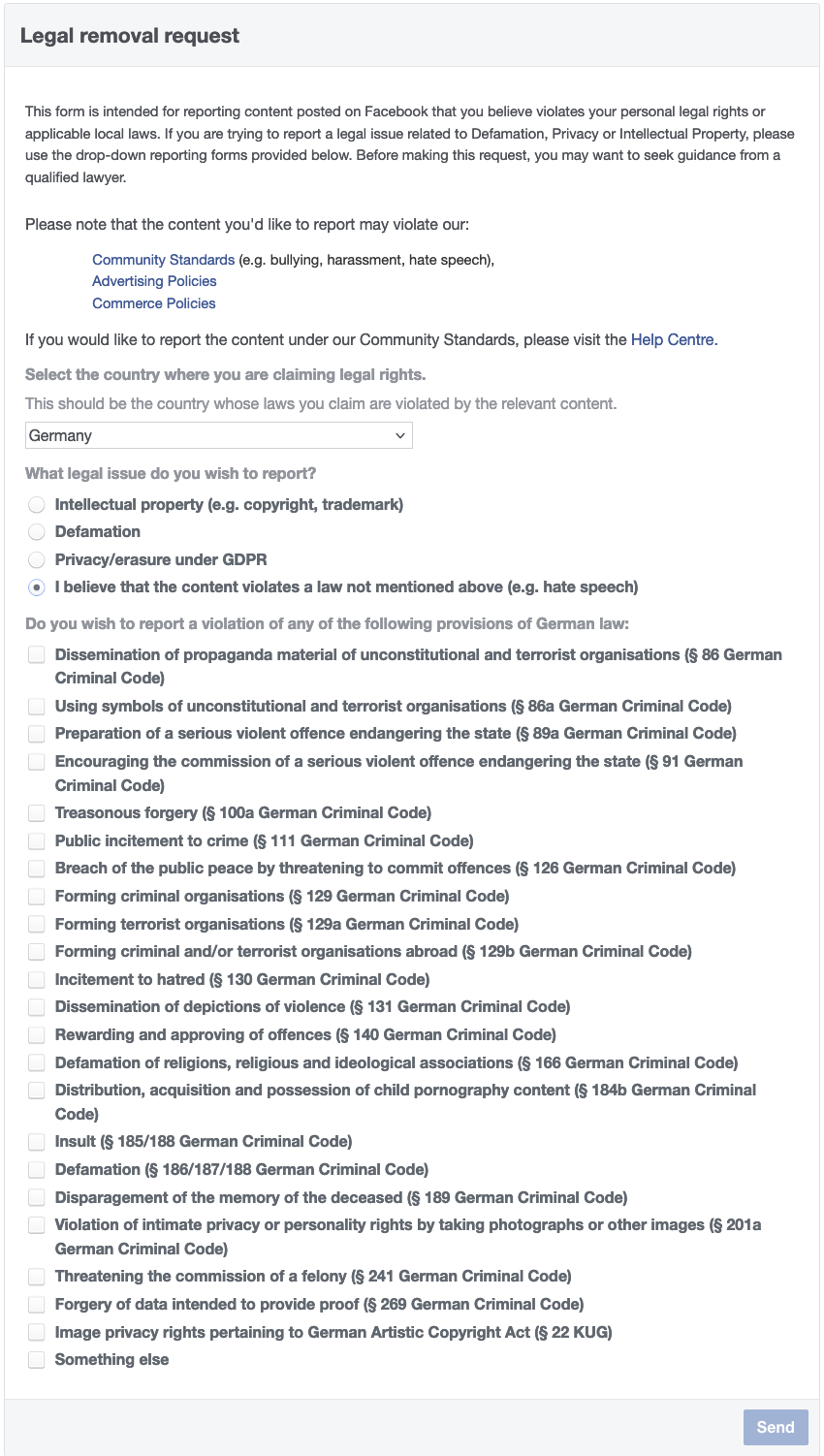}
  \caption{Example of Facebook's illegal content  reporting UI for Germany}
  \label{fig:legal_form_example}
   \Description{This illustration shows Facebook's UI for reporting illegal content. It contains different legal categories as drop down options connected to legal norms. }
\end{figure}

\subsection{Prompting and evaluation strategy}
\label{prompting}

\subsubsection{Prompting}
A robust prompt template is essential, because the quality of an LLM’s answers depends strongly on the phrasing, specificity and structure of the prompt.  Empirical work on \emph{policy–as-prompt} shows that clear, organized prompts, with bullet points and sections rather than long paragraphs, yield higher accuracy than wordy or verbose prompts  and that even small formatting changes can cause different outputs (predictive multiplicity) \cite{palla2025policyaspromptrethinkingcontentmoderation}.  
Prompting as an application of reasoning in the use of LLMs is proposed in utilizing examples or descriptions of desired output in a structured way.
Research on \emph{argumentative LLMs} provides an explicit method for generating \emph{arguments for and against} a decision so that outputs are explainable and contestable \cite{freedman2025argumentativelargelanguagemodels}.

Recent work on ranking arguments recommends structuring prompts into three parts: 
\begin{itemize}
    \item A system message defining the model’s role, 
    \item A task description with clear criteria and numbering, and
    \item Formatting examples that illustrate the desired output \cite{elaraby-etal-2024-persuasiveness}. 
\end{itemize}

The template is designed to generate a \emph{stable set of responses} for each piece of content in the experiment.  It supports two LLM agents—a \textbf{platform-leaning agent} that interprets content through the lens of legal definitions (e.g., the Digital Services Act and other relevant national provisions).

We generated the \emph{Case Descriptions} with ChatGPT 4.1 based on the following prompting template and automatically evaluated the outcome by a judge model (ChatGPT 4.5). This automatic evaluation was in a next step evaluated by two of the authors who have been working in legal adjacent fields for more than four years and have knowledge about German national law: 

\begin{itemize}
\item ChatGPT4.1: Make a short summary of the cases provided in the excel file.
\end{itemize}

\subsubsection{Scenario: Chatgpt 4.1.:}

System Message: You are generating detailed, realistic scenarios for a user study on illegal content flagging in online platforms. For each case, use all available information (summary and full case via link) to write a single, clear scenario of user-generated content or action that would qualify as potentially illegal for the specific category. The scenario should be plausible, anonymized, understandable to a layperson, and sufficiently detailed for use in a user experiment.

\subsubsection{Instructions:}
Read the full case and summary.
Write a stand-alone scenario describing a post, comment, message, or online action.
The scenario should:
Clearly illustrate the type of illegal content (e.g., hate speech, defamation, drug promotion, threats, etc.)
Be specific and detailed enough for users to understand the situation.
Be anonymized (no real names).
Use plain language, without legal or procedural references.
Provide context and any relevant reactions, if they help clarify the situation.
Be up to 200 words if needed for clarity.
The output should be only the scenario text—do not include the case summary, statute, link, or any other information.

\subsubsection{Example Output:} A user posts in a local Facebook group: “Don’t trust the owner of the bakery on Main Street. He poisons his customers and cheats people out of money. Spread the word so nobody else gets scammed!” Several group members reply with supportive comments and share the post. Others ask for evidence, but the original poster does not respond.

Prompt for LLM or Human Writer (per case):
Write a detailed, realistic, and clear scenario describing user-generated content or online behavior that would fit the category of illegal content in a social media context. Use the full case and summary for context and details. Do not reference the case, statute, or any legal terms. Only output the scenario.
\begin{itemize}
\item Judge: Chat GPT 4.5: Take the summary information and check with the case PDF in batches (1-4) if the summaries are correct.
\end{itemize}

\subsection{Reporting scenarios and  legal cases for reporting illegal content in Germany}
\label{tab:cases-appendix}

\begin{table*}[H]
  \caption{Cases: category, law reference, case description, and source (Part 1 of 3)}
  \label{tab:cases-part1}
\scriptsize
  \centering
  \setlength{\tabcolsep}{3pt} 
  \begin{tabular}{@{}%
    >{\raggedright\arraybackslash}p{0.06\textwidth}%
    >{\raggedright\arraybackslash}p{0.19\textwidth}%
    >{\raggedright\arraybackslash}p{0.22\textwidth}%
    >{\raggedright\arraybackslash}p{0.35\textwidth}%
    >{\raggedright\arraybackslash}p{0.05\textwidth}@{}}
    \toprule
    \textbf{ID} & \textbf{Category} & \textbf{Law Reference} & \textbf{Case Description} & \textbf{Source} \\
    \midrule
    L\textunderscore 1 & Copyright & \S~97 German Copyright Act & A user uploads a full-length movie to a public Facebook group without the rights holder’s permission, writing, “Watch this new release for free, no need to pay for streaming!” The post quickly gains attention, with other users commenting their thanks and sharing the video with friends. Some users ask if there are more movies available, and the original poster replies with links to several other recent films they have uploaded. & \cite{noauthor_beschluss_2018}\\
    \midrule
    L\textunderscore 2 & Trademark & \S~14 Trademark Act (MarkG) & A user creates an Instagram account using the name and logo of a well-known sports brand, posting images of clothing and shoes with descriptions like “Official products – limited sale!” The account’s posts direct interested buyers to a private messaging chat, where the user offers counterfeit goods using the brand’s trademarks and claims they are authentic. Several followers ask questions about product authenticity, but the account owner insists all items are genuine. & \cite{noauthor_urteil_2014-1}\\ 
    \midrule
    L\textunderscore 3 & Defamation & \S~187 German Criminal Code & A member of a local Facebook group posts: “Don’t ever trust the manager of the new restaurant on Oak Street. He steals tips from employees and serves expired food. People have gotten sick because of him—share this so nobody goes there!” The post is widely shared in the community. Some users reply with similar accusations, while others defend the restaurant and ask for evidence, which the original poster does not provide. & \cite{noauthor_urteil_2017}\\
    \midrule
    L\textunderscore 4 & GDPR & Art 32(1) \& Recital 75; Art 21 (1); Art 17 (1)(d); Art 16 & An individual posts in a public  group for parents at a local school, uploading a spreadsheet with the names, addresses, phone numbers, and grades of all students in the class. The message reads, “For anyone who wants to arrange playdates or study groups—here’s all the info you need!” Other parents react with concern about the exposure of sensitive information, and some ask for the file to be removed, but the poster argues it is helpful for organizing events. & \cite{noauthor_fashion_2019}\\
    \midrule
    L\textunderscore 5 & Dissemination of propaganda material of unconstitutional and terrorist organizations  & \S~86 German Criminal Code & A user uploads a video to a public online forum showing banners and flyers with the symbols and slogans of a banned extremist group. In the post, the user writes: “Learn the truth about our movement and why the media is lying—download and share these flyers with everyone you know!” Other forum members respond by reposting the video and materials in additional groups, encouraging others to join their cause and spread the propaganda further. & \citep{noauthor_urteil_2005} \\
    \midrule
    L\textunderscore 6 & Using symbols of unconstitutional and terrorist organizations  & \S~86a German Criminal Code & A user uploads a profile picture to a popular online forum that features the emblem of a banned extremist organization. In their profile bio, they include slogans and hashtags supporting the group’s ideology. Other users notice the symbol and some report the account, while a few others comment expressing agreement with the views. & \cite{noauthor_urteil_2022-1}\\
    \midrule
    L\textunderscore 7 & Preparation of a serious violent offense endangering the state  & \S~89a German Criminal Code & In a closed chat group, a user posts detailed instructions on how to make homemade explosives, followed by a message: “Let’s meet on Friday to plan everything—our target is the city’s government office. Download the attached files and be ready.” Members discuss the plan and share suggestions for avoiding detection. & \cite{noauthor_beschluss_2014}\\
    \midrule
    L\textunderscore 8 & Encouraging the commission of a serious violent offense endangering the state  &  \S~91 German Criminal Code & On a public Telegram channel, a user writes: “It’s time to take action against the authorities. If we all attack together, they can’t stop us. Let’s make an example they won’t forget—who’s in?” The post gets replies from several users expressing support and making plans to join in.  & \cite{noauthor_beschluss_2017} \\
    \midrule
    L\textunderscore 9 & Treasonous forgery &  \S~100a German Criminal Code & A user shares a digitally altered government document in a Facebook group, claiming it is a leaked official paper showing new secret laws. The post reads: “Here’s the proof they’re lying to us—share this before it gets deleted!” The fake document is downloaded and spread widely among group members, who begin discussing supposed conspiracies. & \cite{noauthor_endurteil_2020} \\
    \bottomrule
  \end{tabular}
\end{table*}
\begin{table*}[H]
  \caption{Cases: category, law reference, case description, and source (Part 2 of 3)}
  \label{tab:cases-part2}
 \scriptsize
  \centering
  \setlength{\tabcolsep}{3pt}
  \begin{tabular}{@{}%
    >{\raggedright\arraybackslash}p{0.06\textwidth}%
    >{\raggedright\arraybackslash}p{0.19\textwidth}%
    >{\raggedright\arraybackslash}p{0.22\textwidth}%
    >{\raggedright\arraybackslash}p{0.35\textwidth}%
    >{\raggedright\arraybackslash}p{0.05\textwidth}@{}}
    \toprule
    \textbf{ID} & \textbf{Category} & \textbf{Law Reference} & \textbf{Case Description} & \textbf{Source} \\
    \midrule
    L\textunderscore 10 & Public incitement to crime  & \S~111 German Criminal Code & A user comments under a video about environmental protests: “The only way to stop them is to burn down their offices! If you see one of their buildings, set it on fire. Who’s with me?” Several other users react with approval, some sharing tips on how to avoid getting caught. & \cite{noauthor_beschluss_2018}\\
    \midrule
    L\textunderscore 11 & Breach of the public peace by threatening to commit offenses  & \S~126 German Criminal Code & A user creates a series of posts in a large online discussion board, saying: “Tomorrow at noon, I’m going to bring a weapon to the city center and show everyone what happens when you ignore me. Watch out—you’ve all been warned.” The post causes widespread alarm, with many users contacting the police and warning others to avoid the area. & \cite{noauthor_urteil_2020} \\
    \midrule
    L\textunderscore 12 & Forming criminal organizations  & \S~129 German Criminal Code & On an encrypted app, a user creates a group titled “Get Rich Fast Crew” and invites others to join. The group’s description reads: “Looking for people to help with online scams—bank card fraud, phishing, fake stores. Message if interested in making money fast, no questions asked.” New members introduce themselves and discuss division of tasks. & \citep{noauthor_beschluss_2018-1}\\
    \midrule
    L\textunderscore 13 & Forming terrorist organizations &  \S~129a German Criminal Code & A user starts a new channel on an online platform, inviting people to join “The Resistance Force.” The group openly discusses violent plans to attack infrastructure and government buildings, recruits new members, and shares propaganda videos encouraging followers to carry out attacks for their cause. & \cite{noauthor_beschluss_2017-2} \\
    \midrule
    L\textunderscore 14 & Forming criminal and/or terrorist organizations abroad   & \S~129b German Criminal Code & A group of users creates a private forum for planning coordinated attacks on specific targets, sharing maps, blueprints, and lists of supplies needed for each operation. They assign roles (leader, logistics, lookout) and agree on code words for communication. The group regularly recruits new members and celebrates past illegal actions. & \cite{noauthor_beschluss_2021} \\
    \midrule
    L\textunderscore 15 & Incitement to hatred  & \S~130 German Criminal Code & In a public Facebook group, a user posts: “Foreigners are taking over our town and ruining everything. We need to stand up and drive them out before it’s too late. Don’t talk to them, don’t help them—let’s make them feel unwelcome so they leave!” The post sparks a heated debate, with some users supporting the message and others condemning it. & \citep{noauthor_beschluss_2017-3}\\
    \midrule
    L\textunderscore 16 & Dissemination or depictions of violence  & \S~131 German Criminal Code & A user creates a post in a messaging app group, attaching a PDF filled with leaflets that glorify the actions and ideology of a banned extremist organization. The user writes, “Read and spread the truth—don’t let the government silence us!” Other members forward the document to additional groups and praise the message. & \cite{noauthor_beschluss_2020} \\
    \midrule
    L\textunderscore 17 & Rewarding and approving of offences  & \S~140 German Criminal Code & During a livestream on a social media platform, a user displays a large flag in the background featuring the emblem of a prohibited political movement. Throughout the video, they make repeated references to the movement’s slogans and encourage viewers to show their support by displaying similar symbols in public. & \cite{noauthor_beschluss_2017-1} \\
    \midrule
    L\textunderscore 18 & Defamation of religions, religious and ideological associations  & \S~166 German Criminal Code & In the comments section under a viral video, a user writes: “If you see someone from the rival group, don’t hesitate—beat them up and teach them a lesson! If we all act together, they can’t stop us.” Several users reply with agreement and share locations where members of the rival group often gather. & \cite{noauthor_urteil_2021} \\
    \bottomrule
  \end{tabular}
\end{table*}
\begin{table*}[H]
  \caption{Cases: category, law reference, case description, and source (Part 3 of 3)}
  \label{tab:cases-part3}
\scriptsize
  \centering
  \setlength{\tabcolsep}{3pt}
  \begin{tabular}{@{}%
    >{\raggedright\arraybackslash}p{0.06\textwidth}%
    >{\raggedright\arraybackslash}p{0.19\textwidth}%
    >{\raggedright\arraybackslash}p{0.22\textwidth}%
    >{\raggedright\arraybackslash}p{0.43\textwidth}%
    >{\raggedright\arraybackslash}p{0.05\textwidth}@{}}
    \toprule
    \textbf{ID} & \textbf{Category} & \textbf{Law Reference} & \textbf{Case Description} & \textbf{Source} \\
    \midrule
    L\textunderscore 19 & Distribution, acquisition and possession of child pornography  content  & \S~184b German Criminal Code & A user posts on a community forum: “Beware of the dentist on Linden Street—he botches treatments on purpose to make more money from follow-up visits. He’s a fraud and shouldn’t be allowed to practice!” The post quickly gains traction, with others sharing it and some contributing their own negative (but unverified) experiences. & \cite{noauthor_beschluss_2013-1} \\
    \midrule
    L\textunderscore 20 & Insult  & \S~185/188 German Criminal Code & After losing a debate in a public Facebook group, a user replies to their opponent’s comment: “You are the dumbest person here. Everyone is laughing at how stupid your arguments are. Maybe you should just delete your account.” Several other users join in, mocking and insulting the original poster. & \cite{noauthor_urteil_2016} \\
    \midrule
    L\textunderscore 21 & Defamation  & \S~186/187/188 German Criminal Code & A user posts in a neighborhood Facebook group: “Don’t trust the owner of the corner shop. He scams his customers, sells expired products, and cheats people out of their money every day. I heard he was even investigated by the police for fraud!” Group members react by sharing the post, leaving negative comments, and warning others to avoid the shop. Some ask for proof, but the original poster doesn’t respond. & \cite{noauthor_beschluss_2020-1} \\
    \midrule
    L\textunderscore 22 & Disparagement of the memory of the deceased  &  \S~189 German Criminal Code & After a local public figure passes away, a user shares a meme in a public  channel showing the person with text implying they were a criminal and deserved to die. The meme quickly circulates, with some users reacting with laughing emojis and others condemning the post as disrespectful and hateful toward the deceased and their family. & \cite{noauthor_beschluss_2022}\\
    \midrule
    L\textunderscore 23 & Violation of intimate privacy pf personality rights by taking photographs or other images  & \S~201a German Criminal Code & A user secretly takes intimate photos of their ex-partner without consent and uploads them to a private forum, captioning the images: “This is what she’s really like—thought everyone should know.” The post includes personal details and encourages others to share the images further. Several forum members comment, and the images spread to other platforms. & \cite{noauthor_urteil_2020}\\
    \midrule
    L\textunderscore 24 & Forgery of data intended to provide proof &  \S~269 German Criminal Code & During an online dispute, a user uploads a screenshot of an email that appears to be from another group member, supposedly admitting to misconduct. The email is actually a digitally altered forgery created to win the argument and discredit the other person. Other users see the screenshot and begin to attack the accused member based on the falsified evidence.  & \cite{noauthor_beschluss_2018-3}\\
    \midrule
    L\textunderscore 25 & Image privacy rights pertaining to German Artistic Copyright Act  & \S~22 KUG & A user uploads a photograph of a stranger sitting in a park to their public Instagram account, tagging the location and writing: “Look who I spotted—does anyone know this person?” The individual in the photo did not give consent to be photographed or published. The image is widely shared, and the person becomes the subject of unwanted online attention.  & \cite{noauthor_vom_2018} \\
    \midrule
    L\textunderscore 26 & Threatening the commission of a felony  & \S~241 German Criminal Code & A user sends a message in a gaming  server to another player after losing a match: “If you don’t delete your account, I’ll find you and burn your house down. You’ll regret ever messing with me.” Other members of the server see the message and express concern for the recipient’s safety.  & \cite{noauthor_landesrechtsportal_nodate}\\
    \midrule
    L\textunderscore 27 & Something else & (e.g. \S~30a Abs. 1 BtMG) & A user posts on a public Telegram channel: “I’ve got the best quality MDMA and LSD, straight from the source. DM me if you want to buy—safe, fast, and anonymous. Don’t miss out!” Several users send private messages to place orders. & \cite{noauthor_urteil_2023-1}\\
    \bottomrule
  \end{tabular}
\end{table*}

\section{Additional Analyses for H4: Explanation-of-Reasons (EoR) Quality}
\label{app:h4}

\subsection{Inter-Rater Reliability}

Each explanation was independently rated by two trained annotators on four dimensions:
\emph{element coverage}, \emph{proportionality reasoning}, \emph{reasoning depth}, and
\emph{perceived overall quality}, using 1--5 Likert scales.

Inter-rater reliability was acceptable to high across all dimensions.
Krippendorff’s $\alpha$ ranged from 0.67 to 0.82, and quadratic-weighted Cohen’s $\kappa$
ranged from 0.69 to 0.82.
Agreement within $\pm 1$ scale point exceeded 79\% for all dimensions.
These values meet or exceed commonly accepted thresholds for subjective content analysis,
supporting aggregation of coder ratings via the mean.
Full reliability statistics are reported in Table~\ref{tab:h4_irr}.

\begin{table}[H]
\centering
\footnotesize
\caption{Inter-rater reliability for EoR quality dimensions.}
\label{tab:h4_irr}
\begin{tabular}{lcccc}
\toprule
\textbf{Dimension} & $\boldsymbol{\alpha}$ & $\boldsymbol{\kappa_q}$ & Exact Agree. & Agree. $\pm 1$ \\
\midrule
Element coverage & 0.76 & 0.75 & 0.82 & 0.86 \\
Proportionality reasoning & 0.82 & 0.82 & 0.86 & 0.89 \\
Reasoning depth & 0.67 & 0.69 & 0.75 & 0.79 \\
Perceived overall quality & 0.69 & 0.70 & 0.80 & 0.83 \\
\bottomrule
\end{tabular}
\end{table}

\subsection{Model Specification and Robustness}

In the main analysis, we treated coder-mean ratings as approximately continuous and estimated
ordinary least squares (OLS) regression models of the form:
\[
Y_{ij} = \beta_0 + \beta_1\,\mathrm{Manual}_i + \beta_2\,\mathrm{Evaluative}_i + \varepsilon_{ij},
\]
with Conventional XAI as the reference condition.

Because participants could contribute more than one rated report, we used participant-clustered robust standard errors whenever repeated observations were present;
otherwise, HC1 heteroskedasticity-robust standard errors were applied.
All reported inference is two-sided.

As a robustness check, we additionally estimated cumulative logit (ordered logit) models treating
the outcomes as ordinal rather than continuous.
Across all four dimensions, the ordered-logit results were directionally consistent with the OLS
estimates and led to identical substantive conclusions:
no reliable advantage of Evaluative AI over Conventional XAI or Manual reporting.
Ordered-logit model summaries and Wald contrasts are reported in
Appendix Tables~\ref{tab:h4_ols}--\ref{tab:h4_ols_contrasts}.

\begin{table}[H]
\centering
\footnotesize
\caption{OLS regression results for Explanation-of-Reasons (EoR) quality (coder-mean ratings).
Conventional XAI is the reference condition. Participant-clustered robust standard errors.}
\label{tab:h4_ols}
\begin{tabular}{l l r r r r}
\toprule
\textbf{Dimension} & \textbf{Term} & \textbf{Coef.} & \textbf{SE} & \textbf{95\% CI} & \textbf{\emph{p}} \\
\midrule
Element coverage
& Evaluative vs.\ XAI & 0.01 & 0.11 & [$-$0.22, 0.23] & .95 \\
& Manual vs.\ XAI     & $-$0.13 & 0.11 & [$-$0.36, 0.09] & .24 \\

Proportionality reasoning
& Evaluative vs.\ XAI & 0.03 & 0.13 & [$-$0.23, 0.29] & .82 \\
& Manual vs.\ XAI     & $-$0.05 & 0.13 & [$-$0.30, 0.20] & .69 \\

Reasoning depth
& Evaluative vs.\ XAI & 0.07 & 0.12 & [$-$0.16, 0.30] & .56 \\
& Manual vs.\ XAI     & $-$0.10 & 0.12 & [$-$0.33, 0.13] & .39 \\

Perceived overall quality
& Evaluative vs.\ XAI & $-$0.01 & 0.12 & [$-$0.23, 0.22] & .95 \\
& Manual vs.\ XAI     & $-$0.15 & 0.12 & [$-$0.39, 0.09] & .22 \\
\bottomrule
\end{tabular}
\end{table}

\begin{table}[H]
\centering
\footnotesize
\caption{Pairwise contrasts for EoR quality dimensions (coder-mean).
Participant-clustered robust standard errors.}
\label{tab:h4_ols_contrasts}
\begin{tabular}{l l r r r}
\toprule
\textbf{Dimension} & \textbf{Contrast} & \textbf{Diff.} & \textbf{SE} & \textbf{\emph{p}} \\
\midrule
Element coverage
& Evaluative vs.\ XAI & 0.01 & 0.11 & .95 \\
& Evaluative vs.\ Manual & 0.14 & 0.11 & .22 \\

Proportionality reasoning
& Evaluative vs.\ XAI & 0.03 & 0.13 & .82 \\
& Evaluative vs.\ Manual & 0.08 & 0.13 & .53 \\

Reasoning depth
& Evaluative vs.\ XAI & 0.07 & 0.12 & .56 \\
& Evaluative vs.\ Manual & 0.17 & 0.12 & .16 \\

Perceived overall quality
& Evaluative vs.\ XAI & $-$0.01 & 0.12 & .95 \\
& Evaluative vs.\ Manual & 0.06 & 0.12 & .61 \\
\bottomrule
\end{tabular}
\end{table}

\subsection{Codebook}
\begin{table*}[h]
  \label{tab:coding_dimensions_scales}
  \centering
  \footnotesize
  \setlength{\tabcolsep}{4.5pt}
  \renewcommand{\arraystretch}{1.12}

  \begin{tabularx}{\textwidth}{@{} >{\raggedright\arraybackslash}p{0.23\textwidth} X @{}}
    \toprule
    \textbf{Dimension} & \textbf{Definition, item, and scale} \\
    \midrule

    \textbf{Element coverage} &
    \textbf{Definition:} Extent to which participants include all required legal elements in their explanation.\par
    \textbf{Item:} ``The explanation covered all legally relevant elements required for this category.''\par
    \textbf{Scale:}
    \begin{tabular}[t]{@{}l@{}}
      1 = None of the elements covered \\
      2 = Few elements covered (1--2 elements) \\
      3 = About half of the elements covered \\
      4 = Most elements covered (more than half) \\
      5 = All elements covered comprehensively
    \end{tabular}
    \\
    \addlinespace

    \textbf{Proportionality reasoning} &
    \textbf{Definition:} Depth, proportionality, and correctness of reasoning.\par
    \textbf{Item:} ``The explanation was clear, accurate, and legally proportional.''\par
    \textbf{Scale:}
    \begin{tabular}[t]{@{}l@{}}
      1 = Very poor (unclear, irrelevant, incorrect) \\
      2 = Weak (minimal detail, vague, partially incorrect) \\
      3 = Moderate (some relevant detail, but limited accuracy/proportionality) \\
      4 = Good (clear, relevant, mostly correct, proportionate) \\
      5 = Excellent (comprehensive, precise, legally sound, proportionate)
    \end{tabular}
    \\
    \addlinespace

    \textbf{Reasoning depth} &
    \textbf{Definition:} Extent to which a participant's explanation demonstrates depth of thought, elaboration,
    and integration of legal or contextual reasoning.\par
    \textbf{Item:} ``The explanation demonstrates deep reasoning and elaboration.''\par
    \textbf{Scale:}
    \begin{tabular}[t]{@{}l@{}}
      1 = Very shallow (minimal or no reasoning, single-word/phrase) \\
      2 = Shallow (basic reasoning, limited detail, mostly surface-level) \\
      3 = Moderate (some elaboration, partial justification, somewhat developed) \\
      4 = Deep (clear justification with multiple relevant points, good detail) \\
      5 = Very deep (comprehensive, well-developed reasoning with nuanced legal/contextual arguments)
    \end{tabular}
    \\
    \addlinespace

    \textbf{Moderation decision} &
    \textbf{Definition:} Rating of the decision outcome assigned by the participant.\par
    \textbf{Item:} ``The participant selected the appropriate moderation decision for the case.''\par
    \textbf{Scale:} Use the task's decision labels/codes (report separately if nominal; do not force 1--5 unless your instrument does so).
    \\
    \addlinespace

    \textbf{Perceived overall quality} &
    \textbf{Definition:} Rater's perceived overall rating of the explanation.\par
    \textbf{Item:} ``Overall, the explanation was of high quality.''\par
    \textbf{Scale:} 1--5, where 5 = very good.
    \\
    \bottomrule
  \end{tabularx}
    \caption{Codebook used for evaluating participants' explanations of the reason.}
\end{table*}

\section{Appendix: Detailed Analysis Plan}
\label{appendix:analysis_plan}

\subsection{Overview}

This appendix provides the full analysis specifications corresponding to the compressed methodology reported in the main text. It documents outcome definitions, model equations, robustness checks, and supplementary analyses for H1--H6. Unless otherwise noted, all hypothesis tests are two-sided and all confidence intervals are 95\%.

\subsection{Outcome Definitions and Coding}

\textbf{Provision accuracy.}
Provision accuracy $Y^{\mathrm{acc}}_{ij}\in\{0,1\}$ indicates whether participant $i$ selected the ground-truth legal provision for case $j$.

\textbf{Misclassification distance.}
Misclassification distance $D_{ij}\in\{0,1,2,3\}$ captures how far the selected provision deviated from the ground truth, where 0 = correct, 1 = near-miss, 2 = related-but-wrong, and 3 = out-of-scope. This ordinal outcome applies to the participant's selected provision; it should not be confused with the AI error condition itself, which is modeled categorically.

\textbf{Explanation-of-Reasons outcomes.}
Explanation-of-Reasons (EoR) quality was rated independently by two trained coders on four 1--5 dimensions: (i) element coverage, (ii) proportionality reasoning, (iii) reasoning depth, and (iv) perceived overall quality. For the primary analyses, we use the mean of the two coder ratings for each dimension. Reliability statistics are reported separately in the appendix.

\textbf{Downstream risk outcomes.}
For Overbreadth error trials, \emph{over-removal} is a binary indicator equal to 1 if the participant selected a severe enforcement action (e.g., content removal, account suspension, or deplatforming), and 0 otherwise. For Out-of-Scope error trials, \emph{misrouting} is a binary indicator equal to 1 if the participant failed to route the case as out-of-scope, and 0 otherwise.

\textbf{Quality-control exclusions.}
Analyses exclude participants removed for technical issues and, where specified, participants who reported benign filler posts. Excluding participants with false-positive reports on benign filler posts is treated as a robustness check for several hypotheses.

\subsection{H1: Provision Accuracy Under AI Error}

\emph{Estimand.}
Among AI-assisted participants (Evaluative AI vs.\ Conventional XAI), we estimate the effect of Evaluative AI on provision-level accuracy on illegal-content error trials (one error trial per participant by design). Benign posts are excluded; excluding participants who reported benign filler posts is a sensitivity check.

\emph{Outcome.}
Provision accuracy $Y^{\mathrm{acc}}_{ij}\in\{0,1\}$ indicates whether participant $i$ selected the ground-truth provision for case $j$.

\emph{Model.}
We fit a logistic GLM with AI error condition and case fixed effects:
\[
\Pr(Y^{\mathrm{acc}}_{ij}=1)
= \mathrm{logit}^{-1}\!\Big(
\beta_0
+ \beta_1\,\mathrm{Evaluative}_i
+ \boldsymbol{\gamma}^{\top}\mathbf{E}_j
+ \boldsymbol{\eta}^{\top}(\mathrm{Evaluative}_i \times \mathbf{E}_j)
+ \boldsymbol{\delta}^{\top}\mathbf{C}_j
\Big),
\]
where $\mathbf{E}_j$ are AI error condition indicators (NM/OB/OS) and $\mathbf{C}_j$ are case fixed effects. Inference uses HC1 robust standard errors; participant-clustered standard errors are reported as a robustness check (numerically similar given one observation per participant). We report odds ratios with 95\% confidence intervals and two-sided $p$-values.

\emph{Robustness.}
We re-estimate the model with pre-task covariates (MAILS, NCS-6, legal knowledge, propensity to trust automation) and under exclusion of participants who reported benign filler posts. Balance checks by condition and AI error condition accompany Fig.~\ref{fig:h1_accuracy_overall}.

\subsection{H2: Misclassification Distance Under AI Error}

\emph{Estimands and sample.}
On illegal-content error trials in the AI-assisted conditions, H2a tests whether Evaluative AI reduces misclassification distance relative to Conventional XAI, using an ordinal distance $D\in\{0,1,2,3\}$ (0 = correct; 3 = farthest misclassification). H2b tests whether this effect varies by AI error condition, with particular focus on Out-of-Scope errors.

\emph{Model.}
We estimate proportional-odds (ordered logit) models with participant-clustered robust standard errors:
\[
\Pr(D \le k) = \mathrm{logit}^{-1}(\tau_k - \eta), \quad
\eta =
\beta_1\,\mathrm{Evaluative}
+ \boldsymbol{\gamma}^{\top}\mathbf{E}
+ \boldsymbol{\theta}^{\top}(\mathrm{Evaluative} \times \mathbf{E}),
\]
where $\mathbf{E}$ are AI error condition indicators (Overbreadth and Out-of-Scope, with Near-Miss as the reference). Negative coefficients indicate a shift toward smaller misclassification distances. The primary specification excludes case fixed effects; case fixed effects are added only as a design-control robustness check when available.

\emph{Robustness.}
We re-estimate models with pre-task covariates (MAILS, NCS-6, legal knowledge, propensity to trust automation), exclude participants who reported benign filler posts, and conduct non-parametric Mann--Whitney tests (Evaluative $<$ XAI). To isolate Near-Miss vs.\ Overbreadth moderation, we additionally estimate a restricted model with an Evaluative$\times$Overbreadth interaction. Results are summarized in Fig.~\ref{fig:h2a_distance_by_condition}.

\subsection{H3: No-Error Trials (Correct AI Assistance)}

\emph{Panel.}
We analyze no-error trials in the AI-assisted conditions only (Evaluative AI and Conventional XAI) after standard quality control. The panel contains one no-error trial per participant in the AI-assisted conditions (\(N=300\), 150 per condition); the Manual condition did not include no-error trials by design.

\emph{Outcomes and models.}
H3a (accuracy): Provision accuracy is modeled using a binomial GLM with HC1 robust standard errors. Case fixed effects are excluded in the primary specification and added only as a design-control robustness check when available. We additionally conduct a two one-sided tests (TOST) equivalence test on the difference in proportions with margin \(\delta = 0.05\).

H3b (decision time): Decision time (ms) is analyzed on the log scale using OLS with HC1 robust standard errors. Times are winsorized at the 1st/99th percentiles prior to transformation. As a non-parametric check, we report a one-tailed Mann--Whitney U test on raw times testing whether Conventional XAI is faster than Evaluative AI.

\emph{Robustness.}
Results are robust to inclusion of case fixed effects (when available) and to non-parametric decision-time tests.

\subsection{H4: Explanation-of-Reasons Quality}

\emph{Panel.}
We analyze illegal-content reports with completed coder ratings across all three assistance conditions (Manual, Conventional XAI, Evaluative AI). Each report was independently rated by two trained annotators. Participants contributed up to two rated reports, resulting in repeated observations at the participant level.

\emph{Outcomes.}
EoR quality was assessed on four 1--5 rating scales: (i) element coverage, (ii) proportionality reasoning, (iii) reasoning depth, and (iv) perceived overall quality. For each dimension, we use the mean of the two coder ratings as the dependent variable. Inter-rater reliability was acceptable to high across dimensions (Krippendorff’s $\alpha \approx 0.67$--$0.82$; quadratic-weighted Cohen’s $\kappa \approx 0.69$--$0.82$; Appendix), supporting aggregation.

\emph{Model.}
For each outcome, we estimate linear regression models on coder-mean ratings, treating aggregated scores as approximately continuous. Assistance condition is entered as a categorical predictor with Conventional XAI as the reference. Because participants may contribute multiple reports, we use participant-clustered robust standard errors. We report coefficient estimates with 95\% confidence intervals and two-sided $p$-values. Pairwise contrasts (Evaluative vs.\ XAI, Manual vs.\ XAI, Evaluative vs.\ Manual) are evaluated via linear hypothesis tests.

\subsection{H5: Error-Type Impact and Attenuation}

\emph{Panel.}
We restrict analysis to illegal-content trials in the AI-assisted conditions (Conventional XAI and Evaluative AI) and construct a paired difference-in-differences (DiD) panel. Each participant contributes exactly two trials: (i) one illegal-content no-error trial ($\textit{post}=0$) and (ii) one illegal-content error trial ($\textit{post}=1$), where the AI error condition is either Near-Miss, Overbreadth, or Out-of-Scope. AI error condition is therefore assigned between participants, while error presence (correct AI assistance vs.\ AI error condition) varies within participants. We define $\textit{pair\_errorCondition}_i$ as the AI error condition of participant $i$'s error trial and assign it to both rows of the participant’s pair, yielding a constant pair-level factor used for DiD estimation. Benign content is excluded upstream through quality-control filtering.

\emph{Outcomes.}
We analyze two classes of dependent variables. First, provision accuracy is a binary indicator of whether the reported legal provision matches the ground truth. Second, we analyze four coder-rated Explanation-of-Reasons dimensions on 1--5 scales: (i) element coverage, (ii) proportionality reasoning, (iii) reasoning depth, and (iv) perceived overall quality. For each dimension, we compute the mean rating across the two coders. Inter-rater reliability was acceptable across dimensions (Appendix), supporting use of the coder-mean as a stable estimate of explanation quality.

\emph{Models.}
\emph{Provision accuracy (paired DiD GLM).}
We estimate a generalized linear model with a binomial family and logit link on the long-format paired panel, reporting participant-clustered robust standard errors:

\begin{equation}
\begin{aligned}
\Pr(Y^{\mathrm{acc}}_{it}=1)
&= \mathrm{logit}^{-1}\!\Big(
\beta_0
+ \beta_1\,\mathrm{Evaluative}_i
+ \beta_2\,\mathrm{post}_{it} \\
&\quad
+ \boldsymbol{\gamma}^\top \mathbf{E}_i
+ \beta_3\,(\mathrm{Evaluative}_i \times \mathrm{post}_{it}) \\
&\quad
+ \boldsymbol{\eta}^\top (\mathrm{post}_{it} \times \mathbf{E}_i)
+ \boldsymbol{\phi}^\top (\mathrm{Evaluative}_i \times \mathrm{post}_{it} \times \mathbf{E}_i)
\Big).
\end{aligned}
\end{equation}

where $\mathrm{Evaluative}_i$ indicates assistance condition (Evaluative vs.\ Conventional XAI), $\mathrm{post}_{it}$ indicates the error trial ($1$ for error, $0$ for no-error), and $\mathbf{E}_i$ encodes $\textit{pair\_errorCondition}_i$ (Overbreadth and Out-of-Scope, with Near-Miss as the reference category). This paired DiD specification identifies (i) the within-participant accuracy change induced by AI error and (ii) whether that change differs by AI error condition and is attenuated under Evaluative AI.

\emph{Explanation-of-Reasons quality (primary: first-difference OLS).}
Because EoR outcomes are coder-means on 1--5 scales and participants contribute paired observations, our primary analysis uses a participant-level first-difference specification on continuous coder-mean scores. For each participant $i$ and dimension $d$, we compute
\[
\Delta Y^{(d)}_i \;=\; Y^{(d)}_{i,\mathrm{error}} \;-\; Y^{(d)}_{i,\mathrm{noerror}}.
\]
We then estimate OLS models with heteroskedasticity-robust (HC1) standard errors (one row per participant):
\[
\Delta Y^{(d)}_i
=
\alpha_0
+ \alpha_1\,\mathrm{Evaluative}_i
+ \boldsymbol{\kappa}^\top \mathbf{E}_i
+ \boldsymbol{\lambda}^\top (\mathrm{Evaluative}_i \times \mathbf{E}_i)
+ \varepsilon_i.
\]
This removes participant fixed effects by construction and directly captures error-induced changes in explanation quality, while allowing these changes to vary by AI error condition and assistance condition.

\emph{Explanation-of-Reasons quality (sensitivity: ordered logit DiD).}
As a sensitivity analysis, we discretize coder-mean scores back to the original 1--5 categories (rounded to the nearest integer and clipped to $[1,5]$) and estimate proportional-odds (ordered logit) models on the long-format paired panel with participant-clustered robust standard errors. These ordinal models include the same DiD interaction structure and yield substantively consistent conclusions (Appendix).

\emph{Hypothesis tests.}
H5a tests whether the \emph{error penalty} differs by AI error condition. For provision accuracy, we test whether the within-participant error effect (the $\mathrm{post}$ term) varies across $\textit{pair\_errorCondition}$ levels in the Conventional XAI baseline using Wald contrasts (e.g., Out-of-Scope vs.\ Near-Miss; Out-of-Scope vs.\ Overbreadth). For Explanation-of-Reasons quality, H5a is evaluated analogously using Wald contrasts on the AI error condition coefficients in the first-difference models, comparing mean $\Delta Y$ across AI error conditions in the Conventional XAI baseline.

H5b tests attenuation by assistance condition. For provision accuracy, attenuation is assessed via the three-way interaction terms $\mathrm{Evaluative}\times\mathrm{post}\times\textit{pair\_errorCondition}$, with primary focus on the Out-of-Scope attenuation contrast. For Explanation-of-Reasons quality, attenuation is assessed via $\mathrm{Evaluative}\times\textit{pair\_errorCondition}$ interactions in the first-difference models, again with primary focus on Out-of-Scope. We report odds ratios for accuracy models and mean first-difference estimates for Explanation-of-Reasons models, alongside robust confidence intervals and $p$-values. Model-implied margins and descriptive plots are provided to aid interpretation.

\subsection{H6: DSA-Relevant Downstream Risk Outcomes}

\emph{Estimand.}
H6 evaluates whether, relative to Conventional XAI, Evaluative AI leads to user reporting behaviors that are more consistent with downstream risk-relevant outcomes in DSA-aligned reporting, focusing on proportionality and appropriate routing under known AI error conditions.

\emph{Panels and outcomes.}
Analyses are restricted to AI-assisted conditions (Evaluative AI and Conventional XAI) and to trials in which the AI output was erroneous. We estimate separate models by AI error condition. For Overbreadth errors, we operationalize \emph{over-removal} as a binary indicator equal to 1 if the participant selected a severe enforcement action (e.g., content removal, account suspension, or deplatforming), and 0 otherwise. This outcome captures disproportionate intervention in cases where the AI’s error overextends policy scope. For Out-of-Scope (OS) errors, we operationalize \emph{misrouting} as a binary indicator equal to 1 if the participant failed to route the case as out-of-scope, and 0 otherwise. This outcome reflects potential misallocation of moderation effort when content does not fall under the platform’s policy jurisdiction.

\emph{Models.}
For each AI error condition panel, we estimate a binomial generalized linear model with a logit link and participant-clustered standard errors:
\[
\Pr(Y=1) = \mathrm{logit}^{-1}(\alpha + \beta\,\mathrm{Evaluative}),
\]
where $\mathrm{Evaluative}$ is an indicator for Evaluative AI (Conventional XAI is the reference category) and $Y$ denotes either \emph{over-removal} (Overbreadth panel) or \emph{misrouting} (Out-of-Scope panel). We report odds ratios with cluster-robust confidence intervals and present descriptive outcome rates by condition. When outcome variation is insufficient for model estimation, we report descriptive statistics only and treat the corresponding hypothesis test as not identifiable.

\subsection{Participants}

We recruited Prolific participants region-locked to Germany (fluent English; approval rate $\ge 90\%$). Compensation was £1.60 base (10--15 min) plus up to £2.00 bonus (maximum £3.60). We enrolled 489 participants; after exclusions for technical issues or false-positive reports on benign filler posts, $N{=}450$ remained, with 150 participants per condition (\eai{}, \xai{}, and \bl{}) (Appendix Table~\ref{tab:descriptives}). Demographics were 293 men, 152 women, and 5 non-binary participants; mean age was $M{=}31.32$ years ($SD{=}8.7$, range 18--71).

\end{document}